\documentclass[10pt]{article}
\usepackage{appendix}
\usepackage{epsfig}
\usepackage{capt-of}
\usepackage{feynmf}
\usepackage{color}
\usepackage{amsmath,amssymb}

\usepackage[activeacute,english]{babel}
\usepackage{latexsym}

\newcommand{\be}{\begin{equation}}
\newcommand{\ee}{\end{equation}}
\newcommand{\bea}{\begin{eqnarray}}
\newcommand{\eea}{\end{eqnarray}}

\newcommand{\gapp}{\mathrel{\raise.3ex\hbox{$>$}\mkern-14mu
              \lower0.6ex\hbox{$\sim$}}}
\newcommand{\lapp}{\mathrel{\raise.3ex\hbox{$<$}\mkern-14mu
              \lower0.6ex\hbox{$\sim$}}}

\renewcommand\){\right)}
\renewcommand\[{\left[}
\renewcommand\]{\right]}

\newcommand\eq[1]{Eq.~(\ref{#1})}

\newcommand\pa{\partial}

\newcommand\bfk{{\mathbf k}}

\newcommand\bfp{{\mathbf p}}

\newcommand\bfx{{\mathbf x}}

\newcommand\sub[1]{_{\rm #1}}
\newcommand\su[1]{^{\rm #1}}

\newcommand\pz{P_\zeta}
\newcommand\bz{B_\zeta}
\newcommand\tz{T_\zeta}

\newcommand{\no}{\nonumber}
\newcommand{\dn}{\delta N}

\begin{document}

\title{\begin{flushright}
\normalsize PI/UAN-2011-502FT
\end{flushright}
\vspace{5mm}
{\bf Feynman-like Rules for Calculating $n$-Point Correlators of the Primordial Curvature Perturbation}}

\author{
\textbf{C\'esar A. Valenzuela-Toledo$^{1}$\thanks{e-mail: \texttt{cesar.valenzuela@correounivalle.edu.co}}, Yeinzon Rodr\'{\i}guez$^{2,3}$\thanks{e-mail: \texttt{yeinzon.rodriguez@uan.edu.co}}, Juan P. Beltr\'an Almeida$^{2}$\thanks{e-mail: \texttt{juanpbeltran@uan.edu.co}}}\\ \\
\textit{$^{1}$Departamento de F\'isica, Universidad del Valle}  \\
\textit{Ciudad Universitaria Mel\'endez, Santiago de Cali 760032, Colombia} \\ \\
\textit{$^{2}$Centro de Investigaciones, Universidad Antonio Nari\~no}  \\
\textit{Cra 3 Este \# 47A - 15, Bogot\'a D.C. 110231, Colombia} \\ \\
\textit{$^{3}$Escuela de F\'{\i}sica, Universidad Industrial de Santander}  \\
\textit{Ciudad Universitaria, Bucaramanga 680002, Colombia} \\
}

\maketitle

\begin{abstract}
\noindent A diagrammatic approach to calculate $n$-point correlators of the primordial curvature perturbation $\zeta$ was developed a few years ago following the spirit of the Feynman rules in Quantum Field Theory.  The methodology is very useful and time-saving, as it is for the case of the Feynman rules in the particle physics context, but, unfortunately, is not very well known by the cosmology community.  In the present work, we extend such an approach in order to include not only scalar field perturbations as the generators of $\zeta$, but also vector field perturbations.  The purpose is twofold:  first, we would like the diagrammatic approach (which we would call the Feynman-like rules) to become widespread among the cosmology community;  second, we intend to give an easy tool to formulate any correlator of $\zeta$ for those cases that involve vector field perturbations and that, therefore, may generate prolonged stages of anisotropic expansion and/or important levels of statistical anisotropy.  Indeed, the usual way of formulating such correlators, using the Wick's theorem, may become very clutter and time-consuming.
\end{abstract}
\section{Introduction}
Our Universe exhibits departures from the exact isotropy and homogeneity;  we may see it from the distribution of temperatures in the cosmic microwave background radiation (CMB) \cite{wmap7} and from the distribution of matter that gives rise to the large-scale structure \cite{sdss}.  The origin of such departures is a marvelous phenomenon that involves two outstanding mechanisms in particle cosmology \cite{mukhanovbook}:  the production of virtual particles via the Heisenberg uncertainty principle, and the primordial accelerated expansion of the Universe that defines a dynamical particle horizon.  The joint work of these two mechanisms leads to the production of real particles (the Hawking radiation phenomenon) and the classicalisation of the field perturbations that live in the spacetime which describes our Universe. The nature of such field perturbations may be scalar, vectorial, spinorial, or even, it may correspond to p-forms\footnote{The particle production process has been well studied in the scalar field case \cite{mukhanovbook}.  Some nice studies in the vector field case may be found in Refs. \cite{dklr,dimovector,dimosupervector,dkw}.}.  In addition, their contribution to the departure from the exact isotropy and homogeneity (parameterized by the primordial curvature perturbation $\zeta$ \cite{weinbergbook,lythbook}) may be operative during inflation (the standard inflationary mechanism \cite{mukhanovchisov,hawking,starobinsky,guthpi,bardeen,fischler}), at the end of inflation (the inhomogeneous reheating mechanism \cite{dvali}), or after inflation (the curvaton mechanism \cite{lythwands,moroi,lythwands2}).  Depending on the nature of the field perturbations, on the number of fields involved, and on the action describing the theory of gravity and the matter content of the Universe, the statistical properties of the distribution of temperatures in the CMB and of the distribution of matter that form the large-scale structure may exhibit different features: once the variance has been fixed, the power spectrum of $\zeta$ may become scale-dependent, either blue-tilted or red-tilted, or even may become dependent on the direction of the wavevector (signaling violation of the rotational invariance in the two-point correlator of $\zeta$, a property which is called statistical anisotropy \cite{dklr,abramo}).  The higher-order correlators of $\zeta$ are also affected:  the odd-point correlators may become different to zero (signaling departures from a gaussian distribution \cite{lythbook}) and the even-point correlators may become different to products of two-point correlators (signaling also non-gaussianity \cite{lythbook}).  Moreover, the three-point correlator may depend on the directions of the three wavevectors, something which has been called anisotropic non-gaussianity \cite{dimopoulos, Yokoyama:2008xw}. Talking a bit more about the statistical anisotropy, it is generated by either a primordial anisotropic expansion, or a non-scalar nature of the fields involved, or both.

Calculating the $n$-point correlators of $\zeta$ from a well-defined action is crucial in order to compare theory and observation\footnote{At the end, the theory cannot predict the exact outcome of an experiment in a member of the ensemble since the quantum nature only allows us to predict probabilities in the ensemble.  The ergodic theorem is crucial in this respect \cite{weinbergbook} and, therefore, the assumption of statistical homogeneity (translational invariance of the $n$-point correlators) must be maintained.}. The right framework to propagate the statistical properties of the field perturbations to the statistical properties of $\zeta$ is the cosmological perturbation theory (CPT) \cite{weinbergbook,kodamasasaki,mukhanovreport,malikreport}; however, this normally involves lengthy calculations, even more if the nature of the fields is not scalar \cite{himmetoglu4,watanabe2,dulaney,peloso}.  In addition, interesting phenomena such as non-gaussianity may only be accessible if the CPT is taken to second or higher orders \cite{lythbook,lythyeinzon1,yeinzonbook}. Despite of this, the CPT is valid throughout all scales, leaving no room for discrepancies attributed to not considered subhorizon phenomena.  A different approach is via the $\delta N$ formalism \cite{starobinsky2,ss,tanaka,sasaki1}, where $\zeta$ is identified with the perturbation in the number of e-folds of expansion from an initial time in a flat slicing to a final time in a uniform energy density slicing (the threading must be comoving).  The $\delta N$ formalism gives an expression for $\zeta$ which is valid to all orders in CPT;  however, it is only valid for superhorizon scales (in absolut contrast with CPT) \cite{lythbook}.  Of course, extracting the statistical properties of the distribution of $\zeta$ in the $\delta N$ formalism requires also to do some ``perturbation theory'': to expand $\zeta \equiv \delta N$ in a Taylor series and to cut it out at the desired order \cite{yeinzonbook,lythyeinzon2}.  The classicalisation is given at superhorizon scales \cite{lythseery}, as well as the conservation of $\zeta$ is if the adiabatic pressure condition is hold \cite{lythbook,sasaki1};  that is why we do not worry much about the subhorizon scales and still we obtain very precise results at a very cheap computational cost.

The way of calculating $n$-point correlators of $\zeta$ in the $\delta N$ formalism is very similar to the way of calculating scattering amplitudes in the cannonical formulation of Quantum Field Theory \cite{peskin}:  the Wick's theorem is essential \cite{wick}, the calculation is very direct but not so intuitive, and anyway it may become very clutter and time-consuming. Feynman was very clever, even for minor aspects such as avoiding not so intuitive procedures when calculating things;  he developed a series of diagrammatic rules \cite{feynman1,feynman2}\footnote{Actually, this technique was first described by Feynman at the Poconos Conference in 1948, and published in 1951 \cite{feynman3}.} that can reproduce every piece of the calculation of scattering amplitudes, very easy to apply, very easy to remember, and very illuminating at trying to visualize what is actually happening in a quantum process.  Feynman rules became widespread among the particle physicists community and it is nowadays the standard language to interpret quantum processes and calculate scattering amplitudes \cite{diagrammatica}. In view of the above, a few years ago a diagrammatic approach to calculate $n$-point correlators of $\zeta$ was developed including only scalar fields as the generators of $\zeta$ \cite{Byrnes:2007tm}\footnote{A closely related diagrammatic formalism was developed some time ago in order to study large-scale structure formation via gravitational instability; see for instance Ref. \cite{Crocce:2005xy} and references therein. See also Ref. \cite{Giddings:2010ui} for a diagrammatic treatment of perturbative calculations in the ``in-in" formalism and Refs. \cite{Ensslin:2008iu,Ensslin:2010bw} for a diagrammatic approach to information field theory applied to the reconstructions of non-linear signals.}.  Such ``Feynman-like'' rules share the same advantages of its counterpart in Quantum Field Theory, making very easy and intuitive to start any $n$-point correlator calculation.  External lines in this case are identified as Fourier modes of $\zeta$ and internal lines, the propagators, are identified as the power spectra of the scalar field perturbations.  If the field perturbations are non-gaussian, new diagrams appear which do not have a correspondence in the particle physics Feynman rules:  internal lines that split like branches of a tree;  they correspond to connected $n$-point correlators of the field perturbations.  The issue of the vertex renormalization was also treated, showing in an elegant way how to absorb the diagrams with dressed vertices into just one diagram with undressed, but redefined, vertices, in complete analogy with the Feynman rules case \cite{diagrammatica}\footnote{The standard vertex corresponds to a $m$-order derivative (being $m$ the number of internal lines ending in that vertex) of the unperturbed number of e-folds $N$ with respect to the fields involved \cite{Byrnes:2007tm}.}. 

In this paper, we want to draw the attention of the reader into two significant aspects of the history of the Feynman-like diagrammatic approach in cosmology.  First, although very powerful as it is in the particle physics context, not many cosmologists are aware of the existence or the power of the Feynman-like rules, and still most of the calculations based in the $\delta N$ formalism are performed directly, invoking the Wick's theorem.  Of course, the arrival to the relevant integrals is lengthy, the intuitive connection between an important feature in the calculation and the physics behind it is almost lost, and the scene quickly becomes very clutter.  Second, vector field perturbations were not considered in the original formulation of the Feynman-like rules.  Vector fields as possible generators of $\zeta$ have attracted the attention of several scientists as they may generate significant statistical anisotropy, still in agreement with present observations \cite{groeneboom,hanson,hanson2,ma}, via the anisotropic expansion they may generate \cite{himmetoglu4,watanabe2,dulaney,peloso,watanabe1} and also via its own vector nature \cite{dklr,dimovector,dimosupervector,dkw,jabbari1,jabbari2,kostascurvaton2,kostasattractive,karciauskas}.  Indeed, statistical anisotropy surely will become discriminator of models for the generation of $\zeta$ taking into account the forthcoming increment in the precision of observations starting with the PLANCK satellite data \cite{abramo,pullen}.  The first purpose of this paper is to try to disseminate the diagrammatic approach to cosmologists as much as possible; time will tell if this objective was reached.  The second purpose of this paper is to extend the Feynman-like rules to the case where the contributions to $\zeta$ are due to several scalar and vector fields, producing the latter, in general, anisotropic expansion, statistical anisotropy, and anisotropic non-gaussianity \cite{dimopoulos,vrl,vr,mythesis,bartolo1,bartolo2,bartolo3}.  We try to be as general as possible, allowing for an arbitrary number of fields, taking into account the transverse as well as the longitudinal polarizations of the vector fields (i.e. allowing for mass terms), keeping in mind that the action may not be parity-invariant, considering, in general, anisotropic expansion, and assuming that the probability distribution functions for the scalar field perturbations, as well as for the vector field perturbations\footnote{Actually, the probability distribution functions for the scalar perturbations that multiply the respective polarization vectors.}, may in general be non-gaussian.  It is worth emphasizing that, in the context of nowadays cosmology, it is very important to have an efficient method for the calculation of loop corrections; this is because high precision cosmological measurements could be able to access observational signatures encoded in the loop diagrams if they turn to be non-negligible \cite{Sugiyama:2011jt}, so it is crucial to go beyond tree level terms. Moreover, from the theoretical point of view, there exist scenarios in which loop contributions are larger than the tree-level terms and, therefore, are able to produce large non-gaussianities \cite{vrl, vr, Boubekeur:2005fj, Cogollo:2008bi, Rodriguez:2008hy, Kumar:2009ge}; in those scenarios, it is essential to have an efficient method to evaluate  such contributions.

The layout of the paper is as follows.  In Section 2 the $n$-point correlators are introduced;  the meaning of statistical homogeneity, statistical isotropy, and gaussianity is clearly established.  In Section 3 the $\delta N$ formalism is introduced, allowing for the possibility of anisotropic expansion and, therefore, for the existence of vector fields as generators of $\zeta$;  the two-point correlator of $\zeta$ is calculated using the non-diagrammatic technique (i.e. by employing the Wick's theorem) up to the one-loop level.  Section 4 is devoted to the presentation of the Feynman-like rules in the limit where the field perturbations are gaussian; the calculation of the two- and three-point correlator of $\zeta$ up to the one-loop level, and four-point correlator up to the tree level, are performed employing the diagrammatic approach and compared with the findings in Section 3.  Section 5 introduces the rules in the non-gaussian case; again, some examples are considered and compared with the non-diagrammatic approach.  In Section 6 the vertex renormalisation is studied for the gaussian and non-gaussian cases.  The conclusions of this work are presented in Section 7.

\section{Statistical homogeneity, statistical isotropy, and gaussianity}\label{ani-gauss}                 
Quantum mechanics only allows us to predict probabilities of different outcomes after an experiment in an ensemble of systems, in contrast to classical mechanics which does allow us to predict the exact outcome after an experiment in just one element of the ensemble.  Since the underlying physical mechanism in the generation of cosmological perturbations is of quantum nature, the cosmologists are more interested in studying the statistical properties of a perturbation map, say the CMB map or the galaxy distribution map. The way of doing this is via the $n$-point correlators of the perturbations in real space.  Let's define a scalar cosmological perturbation $\beta({\bf x})$ in real space and make a Fourier integral expansion
\begin{equation}
\beta({\bf x}) \equiv \int \frac{d^{3} k}{(2 \pi)^{3}} e^{i {\bf k \cdot x}}\beta({\bf k}) \,, \label{ffe}
\end{equation}
where $\beta({\bf k})$ is the Fourier mode function of $\beta({\bf x})$. The $n$-point correlators of $\beta({\bf x})$ are averages over the ensemble of the products $\beta({\bfx_1}) \beta({\bfx_2}) ... \beta({\bfx_n})$ where ${\bfx_1}$, ${\bfx_2}$, ..., ${\bfx_n}$ represent different points in space\footnote{The ensemble average inside the integral is over the Fourier mode functions only since they are the stochastic variables.}:
\begin{equation}
\langle \beta({\bfx_1}) \beta({\bfx_2}) ... \beta({\bfx_n}) \rangle \equiv \int \frac{d^{3} k_1}{(2 \pi)^{3}} \frac{d^{3} k_2}{(2 \pi)^{3}} ... \frac{d^{3} k_n}{(2 \pi)^{3}} e^{i ({\bfk_1 \cdot \bfx_1} + {\bfk_2 \cdot \bfx_2} + ... + {\bfk_n \cdot \bfx_n})} \langle \beta({\bfk_1}) \beta({\bfk_2}) ... \beta({\bfk_n}) \rangle \,. \label{expan}
\end{equation}
Thus, the correlation functions in real space may be studied via the correlation functions in momentum space.  Let's see now the meaning of statistical homogeneity, statistical isotropy, and gaussianity.

\subsection{Statistical homogeneity}
Of course the perturbation map is not homogeneous (i.e., it is not invariant under spatial translations), but it may be that the probability distribution function governing $\beta({\bf x})$ is, which is called statistical homogeneity \cite{weinbergbook,lythbook,abramo}. This means that the $n$-point correlators in real space are invariant under translations in space, i.e.
\begin{equation}
\langle \beta({\bfx_1} + {\bf d}) \beta({\bfx_2} + {\bf d}) ... \beta({\bfx_n} + {\bf d}) \rangle = \langle \beta({\bfx_1}) \beta({\bfx_2}) ... \beta({\bfx_n}) \rangle \,,
\end{equation}
where ${\bf d}$ is some vector in real space establishing the amount of spatial translation. The only way of achieving this, in view of Eq. (\ref{expan}), is expressing the argument in the exponential function inside the integral as the addition of several terms of the form $f({\bf x}_i - {\bf x}_j)$, which in turn is possible (but it is not the only possibility) if the $n$-point correlators in momentum space are proportional to a Dirac delta function:
\begin{equation}
\langle \beta({\bfk_1}) \beta({\bfk_2}) ... \beta({\bfk_n}) \rangle \equiv (2\pi)^3 \delta^{3}({\bfk_{12...n}}) M_\beta ({\bfk_1}, {\bfk_2}, ... , {\bfk_n}) \,. \label{shcond}
\end{equation}
In the previous expression, ${\bfk_{12...n}}$ means ${\bfk_1} + {\bfk_2} + ... + {\bfk_n}$, and the function $M_\beta ({\bfk_1}, {\bfk_2}, ... , {\bfk_n})$ is called the $(n-1)$-spectrum.  Statistical homogeneity is absolutely necessary as an hypothesis of the ergodic theorem \cite{weinbergbook}; otherwise, although we do have a theoretical framework to do calculations and although we do have a significant amount of observational data with unprecedented precision, we could not compare one with the other.

\subsection{Statistical isotropy}
Once statistical homogeneity has been secured, in the form of Eq. (\ref{shcond}), we ask about the invariance under spatial rotations (i.e. isotropy).  Of course again, the perturbation map is not isotropic, but it may be that the probability distribution function governing $\beta({\bf x})$ is, which is called statistical isotropy \cite{weinbergbook,lythbook,abramo}. This means that the $n$-point correlators in real space are invariant under rotations in space, i.e.
\begin{equation}
\langle \beta(\tilde{\bfx}_1) \beta(\tilde{\bfx}_2) ... \beta(\tilde{\bfx}_n) \rangle = \langle \beta({\bfx_1}) \beta({\bfx_2}) ... \beta({\bfx_n}) \rangle \,,
\end{equation}
where ${\bf \tilde{x}}_i = \mathcal{R} \ {\bf x}_i$, $\mathcal{R}$ being a rotation operator. To satisfy the above requirement, the $(n-1)$-spectrum must satisfy the condition
\begin{equation}
M_\beta (\tilde{\bfk}_1, \tilde{\bfk}_2, ... , \tilde{\bfk}_n) = M_\beta ({\bfk_1}, {\bfk_2}, ... , {\bfk_n}) \,, \label{sicond}
\end{equation}
where the tildes over the momenta represent as well a spatial rotation, parameterized by $\mathcal{R}$, in momentum space. This condition has more explicit consequences in the spectrum (1-spectrum) and the bispectrum (2-spectrum):
\begin{eqnarray}
M_\beta ({\bfk_1}, {\bfk_2}) &\equiv & P_\beta ({\bfk_1}, {\bfk_2}) = P_\beta (k) \,, \label{sicond1} \\
M_\beta ({\bfk_1}, {\bfk_2}, {\bfk_3}) &\equiv & B_\beta ({\bfk_1}, {\bfk_2}, {\bfk_3}) = B_\beta (k_1,k_2,k_3) \,, \label{sicond2}
\end{eqnarray}
where in the first line $k = |{\bfk_1}| = |{\bfk_2}|$, and in the second line $k_i = |{\bfk}_i|$. Starting from the trispectrum (3-spectrum), the condition in Eq. (\ref{sicond}) about statistical isotropy in all the higher-order $(n-1)$-spectra cannot be reduced to similar conditions to the ones in Eqs. (\ref{sicond1}) and (\ref{sicond2}), so that the minimal way of parameterizing the $(n-1)$-spectra (with $n \geq 4$) will always be in terms of all the $n$ wavevectors. The scalar nature of $\beta({\bf x})$ is very important since, if it were a vector or a tensor, there would not be a way to make the $n$-point correlators in real space invariant under spatial rotations. In those cases, we relax the meaning of statistical isotropy and establish that it is present if the $(n-1)$-spectra of the scalar pertubations that multiply the respective polarization vectors or tensors satisfy Eq. (\ref{sicond}).

\subsection{Gaussianity}
Gaussianity may be defined either via the perturbations $\beta({\bf x})$ in real space or via the perturbations $\beta({\bf k})$ in momentum space. Let's talk first about the latter and we will come back later on the former.  We say that the probability distribution function governing $\beta({\bf k})$ is gaussian if for different wavevectors the perturbations are uncorrelated:\footnote{For equal wavevectors, the reality condition on $\beta({\bf x})$ applies and, therefore, there is self-correlation.}
\begin{equation}
\langle \beta({\bfk_1}) \beta({\bfk_2}) \rangle = (2 \pi)^{3} \delta^{3}(\bfk_{12}) P_\beta ({\bfk_1}) \,, \label{2cor}
\end{equation}
and if the $n$-point correlators, with $n$ being odd, are zero, while those with $n$ being even are equal to the sum over all ways of pairing $\beta({\bf k})$s with each other of a product of the 2-point correlators of the pairs\footnote{The sum over pairings does not distinguish those which interchange wavectors in a pair, or which merely interchange pairs.} \cite{weinbergbook,lythbook}:
\begin{eqnarray}
\langle \beta({\bfk_1}) \beta({\bfk_2}) \beta({\bfk_3}) \rangle &=& 0 \,, \label{3cor} \\
\langle \beta({\bfk_1}) \beta({\bfk_2}) \beta({\bfk_3}) \beta({\bfk_4}) \rangle &=&  \langle \beta({\bfk_1}) \beta({\bfk_2}) \rangle \langle \beta({\bfk_3}) \beta({\bfk_4}) \rangle + \langle \beta({\bfk_1}) \beta({\bfk_3}) \rangle \langle \beta({\bfk_2}) \beta({\bfk_4}) \rangle + \nonumber \\
&&  \langle \beta({\bfk_1}) \beta({\bfk_4}) \rangle \langle \beta({\bfk_2}) \beta({\bfk_3}) \rangle  \label{4cor} \\
&=& (2 \pi^{6}) \delta^{3}(\bfk_{12}) \delta^{3}({\bfk_{34}}) P_\beta ({\bfk_1}) P_\beta ({\bfk_3}) + {\rm two \ permutations} \,, \label{gsh}
\end{eqnarray}
and so on. Eqs. (\ref{2cor}) and (\ref{gsh}) clearly show that statistical homogeneity is a necessary but not sufficient condition if the probability distribution function governing $\beta({\bf k})$ is to be gaussian \cite{dklr,lythbook}. Now, coming back to the perturbations $\beta({\bf x})$ in real space, it is possible to show that the gaussianity condition expressed above implies that the probability distribution function $P(\beta({\bf x}))$ is given by
\begin{equation}
P(\beta({\bf x})) = \frac{1}{\sqrt{2 \pi \langle \beta^{2}({\bf x}) \rangle}} e^{-\beta^{2}({\bf x}) / 2 \langle \beta^{2}({\bf x}) \rangle} \label{pdf} \,,
\end{equation}
which is the usual definition of a gaussian probability distribution function.  However, requiring Eqs. (\ref{3cor}), (\ref{4cor}), and so on for $\beta({\bf x})$, instead of $\beta({\bf k})$, also results in Eq. (\ref{pdf}) without requiring gaussianity in the mode functions $\beta({\bf k})$.  In other words, gaussianity in $\beta({\bf k})$ is a sufficient but not necessary condition for gaussianity in $\beta({\bf x})$; indeed, via the central limit theorem \cite{karlin,adler}, $\beta({\bf x})$ may become gaussian just by being expressed as a sum of uncorrelated quantities ($\beta({\bf k})$) even if they are not gaussian.

When the perturbations $\beta({\bf k})$ are non-gaussian, but still assuming statistical homogeneity, the $n$-point correlators in momentum space, with $n \geq 3$, are expressed in terms of ``connected'' $n$-point correlators (identified with a subscript $c$) which establish the departure from the gaussianity condition:
\begin{eqnarray}
\langle \beta({\bfk_1}) \beta({\bfk_2}) \rangle &=& (2 \pi)^{3} \delta^{3}({\bfk_{12}}) P_\beta ({\bfk_1}) \,, \label{ngsp} \\
\langle \beta({\bfk_1}) \beta({\bfk_2}) \beta({\bfk_3}) \rangle &=&  \langle \beta({\bfk_1}) \beta({\bfk_2}) \beta({\bfk_3}) \rangle_c  \label{ngbs} \\
&=& (2 \pi)^{3} \delta^{3}({\bfk_{123}}) B_\beta ({\bfk_1}, {\bfk_2}, {\bfk_3}) \,, \\
\langle \beta({\bfk_1}) \beta({\bfk_2}) \beta({\bfk_3}) \beta({\bfk_4}) \rangle &=&  \langle \beta({\bfk_1}) \beta({\bfk_2}) \beta({\bfk_3}) \beta({\bfk_4}) \rangle_c + \langle \beta({\bfk_1}) \beta({\bfk_2}) \rangle \langle \beta({\bfk_3}) \beta({\bfk_4}) \rangle + \nonumber \\
&&  \langle \beta({\bfk_1}) \beta({\bfk_3}) \rangle \langle \beta({\bfk_2}) \beta({\bfk_4}) \rangle + \langle \beta({\bfk_1}) \beta({\bfk_4}) \rangle \langle \beta({\bfk_2}) \beta({\bfk_3}) \rangle  \label{ngts} \\
&=& (2 \pi)^{3} \delta^{3}({\bfk_{1234}}) T_\beta ({\bfk_1}, {\bfk_2}, {\bfk_3}, {\bfk_4}) + \nonumber \\
&&  (2 \pi^{6}) \delta^{3}({\bfk_{12}}) \delta^{3}({\bfk_{34}}) P_\beta ({\bfk_1}) P_\beta ({\bfk_3}) + {\rm two \ permutations} \,,
\end{eqnarray}
and so on. In the above, $B_\beta ({\bfk_1}, {\bfk_2}, {\bfk_3})$ and $T_\beta ({\bfk_1}, {\bfk_2}, {\bfk_3}, {\bfk_4})$ are called the connected bispectrum and trispectrum of $\beta$.

\section{The primordial curvature perturbation and the $\dn$ formalism}
The $\dn$ formalism \cite{starobinsky2,ss,tanaka,sasaki1} provides a powerful method to evaluate the primordial curvature perturbation $\zeta ({\bf x}, t)$ in terms of the perturbations of the fields at the the time of horizon crossing $t_{*}$ (corresponding to a flat slicing), and the derivatives of the unperturbed number of $e$-foldings $N(t, t_{*})=\int_{t_{*}}^t  H(t')dt'$ with respect to the unperturbed fields evaluated at $t_{*}$.

According to this formalism, once the separate universe approach has been invoked (see also Ref. \cite{wmll}), and a comoving threading has been established, the value of $\zeta$ in a uniform energy density hypersurface at the final time $t$ is given by the perturbation in the time integral of the local volume expansion rate $\theta$ along a curve starting at an initial flat hypersurface at the time $t_{i}$:
\be
\zeta(\bfx,t) \equiv \delta N (\bfx,t, t_i) - \langle \delta N (\bfx,t, t_i) \rangle \,.
\ee
Here, the bracket notation means a ensemble average (which corresponds to a spatial average is there is statistical homogeneity). In many inflationary scenarios, the number $N$ of $e$-foldings depends only on the values of the fields at horizon crossing so we can write the curvature perturbation as an expansion in the perturbations of the fields at this time \cite{lythyeinzon2}. Supposing that inflation is driven by a single scalar field $\phi(\bfx, t)$ which can be decomposed as $\phi = \phi_0 + \delta \phi$ where $\delta \phi$ is the field perturbation at the initial time $t_i = t_*$, the $\dn$ formula acquires the form:
\be
\zeta(\bfx,t) =  \frac{\partial N(t, t_{*})}{\partial \phi (t_{*})} \delta \phi (\bfx, t_{*} ) +  \frac{1}{2!}\frac{\partial^2 N(t, t_{*})}{\partial \phi^2 (t_{*})} \delta \phi^2 (\bfx, t_{*} ) + \frac{1}{3!}\frac{\partial^3 N(t, t_{*})}{\partial \phi^3 (t_{*})} \delta \phi^3 (\bfx, t_{*} ) + \cdots - (\rm{spatial\; average}) \,.
\ee
By doing this expansion, we are neglecting any additional dependence on $\dot{\phi}$ after horizon crossing which implies that we assume that the field  perturbations are strongly suppressed on large scales; this assumption is valid during a slow-roll regime but, in general, it is valid for any light fields present during inflation. 
To simplify the writing of the expressions, in the following we omit the dependence of the field perturbations on the initial time $t_*$; we also omit the spatial average in the $\dn$ formula.

Due to the versatility of the method, it  has been implemented in many different scenarios; for instance, if we assume that the primordial curvature perturbation is generated by $n$ scalar fields, the $\dn$ formalism gives us the following expansion for $\zeta$:
\be
\zeta(\bfx,t)\equiv \delta N(\phi_1(\bfx),\ldots,\phi_n(\bfx),t) = N_I \delta\phi_I(\bfx) + \frac12 N_{IJ}
\delta\phi_I(\bfx) \delta\phi_J (\bfx) + \frac{1}{3!} N_{IJK}
\delta\phi_I(\bfx) \delta\phi_J (\bfx)\delta\phi_K (\bfx) + \ldots \,, \label{dNsc}
\ee
where $N_I\equiv \pa N/\pa \phi_I$, etc. $I = 1, 2, \cdots, n$ and we understand that repeated indices are summed over. In a recent work this formalism was extended to include also vector field perturbations \cite{dklr} (see also Ref. \cite{Yokoyama:2008xw}), and it was shown that in the simplest case where $\zeta$ is generated by a single vector field and a single scalar, the curvature perturbation can be calculated by means of the following expression:
\be
\zeta(\bfx , t)\equiv\delta N (\phi(\bfx),A_i(\bfx),t)=N_\phi \delta\phi + N_i\delta A_i+\frac{1}{2}N_{\phi\phi}
(\delta\phi)^2+
N_{\phi i}\delta\phi\delta A_i+\frac{1}{2}N_{ij}\delta A_i \delta A_j + \ldots \,,\label{dScsvec}
\ee
where
\be
N_\phi\equiv\frac{\partial N}{\partial \phi}\,,\quad
N_{\phi\phi} \equiv\frac{\partial^2 N}{\partial \phi^2}\,,\quad
\quad N_{\phi i}\equiv\frac{\partial^2 N}{\partial A_i\partial\phi}\,, \quad \rm{etc}. \,,
\ee
are the derivatives with respect to the scalar $\phi$ and the spatial components of the vector field ${\bf A}$. It is very important to warning the reader that the extension of the $\delta N$ formalism to the vector field perturbations case requires relaxing one assumption in the separate universe approach: the isotropic expansion.  The consequence of this is to have a tensor perturbation $h_{ij}$ which is time-dependent \cite{dklr}; however, the identification of $\zeta$ with $\delta N$ remains the same no matter that $h_{ij}$ is time-dependent or not. If for some reason the expansion is isotropic, the perturbations in the scalar and vector fields may be statistically isotropic; however, $\zeta$ would not be in general statistically isotropic due to the vector nature of the fields involved (except, of course, for $\phi$). On the contrary, if the expansion is anisotropic, all the perturbations would be automatically statistically anisotropic and, therefore, $\zeta$ would be too;  this would affect also the $N$ derivatives since the background would not be of the Friedmann-Robertson-Walker type anymore\footnote{$N$ depends on the values of the relevant fields at the initial time $t_i$ and on the total (and uniform) energy density $\rho$ at the final time $t$; there is no explicit dependence on the position. The $N$ derivatives are evaluated in the background so there is neither explicit nor implicit dependence on the position unless the background metric is inhomogeneous or some very specific configuration of inhomogeneous background fields renders the background metric homogeneous; however, in the former case, there would be statistical inhomogeneity, making impossible to compare theory and observations. \label{nderx}}.


Now, we move to the more general case in which the inflationary dynamics is driven by multiple scalar and multiple vector fields. We will consider $n$ scalar fields and $m$ vector fields. To deal with the contributions coming from the different fields involved, we introduce the notation:
\be \label{multisv}
\delta \Phi_{\bar{A}} \equiv \left\{ \delta \phi_{I} \, ,\, \delta A_{i}^a \right\} \,.
\ee
The index $\bar{A}$ is separated in two sets, a set of indices $I$ labelling the scalar fields which runs from 1 to $n$ and another set of indices $a$ labelling vector fields which runs from 1 to $m$. The index $i$ specifies the component of any vector field and it runs from 1 to 3 \footnote{This is because, as inflation homogenizes the vector fields ($\partial_i A_\mu^{a} = 0$), we expect the temporal components of them to vanish \cite{dimovector}. Besides, if the vector fields are massless, we can set the temporal components to zero by a gauge choice.}. We use the index $a$ as a supra-index and the index $i$ as a sub-index. Accordingly, the derivatives of $N$ with respect to the fields are separated as follows:
\bea
N_{\bar{A}} &=& \left\{ N_{I} \, , \, N_{i}^a \right\} \,, \\
N_{\bar{A}\bar{B}} &=& \left\{ N_{IJ} \, , \, N_{Ij}^{\;\; b}  \, , \, N_{ij}^{ab}  \right\} \,,
\eea
and so on. For instance, in the notation above, we represent the mixed second derivative with respect to $\phi_I$ and $A_{j}^{b}$ as $N_{Ij}^{\;\; b} \equiv \partial^2 N / \partial \phi_I \partial A_{j}^{b}$. In terms of the notation in Eq. (\ref{multisv}), the curvature perturbation for the multi-scalar and multi-vector field case is written as
\bea
\zeta (\bfx, t) &\equiv & \delta N(\Phi_{\bar{A}}(\bfx), t) \nonumber \\
&=& N_{\bar{A}} \delta \Phi_{\bar{A}}  + \frac{1}{2}N_{\bar{A} \bar{B}} \delta \Phi_{\bar{A}} \delta  \Phi_{\bar{B}}+\frac{1}{3!}N_{ \bar{A} \bar{B} \bar{C}} \delta \Phi_{\bar{A}} \delta  \Phi_{\bar{B}} \delta  \Phi_{\bar{C}} + \frac{1}{4!}N_{  \bar{A} \bar{B} \bar{C} \bar{D}} \delta \Phi_{\bar{A}} \delta  \Phi_{\bar{B}} \delta  \Phi_{\bar{C}} \delta  \Phi_{\bar{D}}+ \cdots \,. \label{zetamultisv}
\eea

 As we can see from the equations above, since $\zeta$ is a series in the field perturbations $\delta\Phi_{\bar{A}}$, then its correlation functions will be expressed as perturbative series in the correlation functions of $\delta\Phi_{\bar{A}}$.
To this end, we first write the mode function associated to $\zeta$ in terms of the mode functions associated to the field perturbations\footnote{Here we take into account Footnote \ref{nderx}: even for those cases where the background metric is anisotropic, but still homogeneous, the $N$ derivatives evaluated in the background do not depend either explicitly nor implicitly on the position.}:
\begin{eqnarray}
\zeta({\bfk}, t) &=& N_{\bar{A}} \delta \Phi_{\bar{A}} ({\bfk}) + \frac{1}{2} N_{\bar{A} \bar{B}} \int \frac{d^{3} k_1}{(2 \pi)^{3}} \delta \Phi_{\bar{A}} ({\bf k} - {\bfk_1}) \delta \Phi_{\bar{B}} ({\bfk_1}) + \nonumber \\
&&  \frac{1}{3!} N_{\bar{A} \bar{B} \bar{C}} \int \frac{d^{3} k_1 d^{3} k_2}{(2 \pi)^{6}} \delta \Phi_{\bar{A}} ({\bf k} - {\bfk_1} - {\bfk_2}) \delta \Phi_{\bar{B}} ({\bfk_1}) \delta \Phi_{\bar{C}} ({\bfk_2}) + ... - \nonumber \\
&&  \frac{1}{2} N_{\bar{A} \bar{B}} (2 \pi)^{3} \delta^{3} ({\bf k}) \langle \delta \Phi_{\bar{A}} ({\bf x}) \delta \Phi_{\bar{B}} ({\bf x}) \rangle - \frac{1}{3!} N_{\bar{A} \bar{B} \bar{C}} (2 \pi)^{3} \delta^{3} ({\bf k}) \langle \delta \Phi_{\bar{A}} ({\bf x}) \delta \Phi_{\bar{B}} ({\bf x}) \delta \Phi_{\bar{C}} ({\bf x}) \rangle - ... \,,
\end{eqnarray}
where in the last line the mode function of the spatial average in the $\delta N$ formula is shown.
Now, we may multiply $n$ mode functions of $\zeta$ and perform the average over the ensemble in order to obtain the $n$-point correlator.  As an example, we will obtain the 2-point correlator of $\zeta$ by truncating the series up to third order and taking into account terms that will only be represented by tree level or one-loop diagrams\footnote{Except for four terms which are formally represented by two-loop disconnected diagrams and that, as we will see, will cancel each other out.}:
\begin{eqnarray}
\langle \zeta({\bfk_1}) \zeta({\bfk_2}) \rangle &=& N_{\bar{A}} N_{\bar{B}} \langle \delta \Phi_{\bar{A}} ({\bfk_1}) \delta \Phi_{\bar{B}} ({\bfk_2}) \rangle + \nonumber \\
&&  \frac{1}{2} N_{\bar{A}} N_{\bar{B} \bar{C}} \int \frac{d^{3} k_3}{(2 \pi)^{3}} \langle \delta \Phi_{\bar{A}} ({\bfk_1}) \delta \Phi_{\bar{B}} ({\bfk_2} - {\bfk_3}) \delta \Phi_{\bar{C}} ({\bfk_3}) \rangle + \nonumber \\
&&  \frac{1}{3!} N_{\bar{A}} N_{\bar{B} \bar{C} \bar{D}} \int \frac{d^{3} k_3 d^{3} k_4}{(2 \pi)^{6}} \langle \delta \Phi_{\bar{A}} ({\bfk_1}) \delta \Phi_{\bar{B}} ({\bfk_2} - {\bfk_3} - {\bfk_4}) \delta \Phi_{\bar{C}} ({\bfk_3}) \delta \Phi_{\bar{D}} ({\bfk_4}) \rangle + \nonumber \\
&&  \frac{1}{2} N_{\bar{A}} N_{\bar{B} \bar{C}} \int \frac{d^{3} k_3}{(2 \pi)^{3}} \langle \delta \Phi_{\bar{A}} ({\bfk_2}) \delta \Phi_{\bar{B}} ({\bfk_1} - {\bfk_3}) \delta \Phi_{\bar{C}} ({\bfk_3}) \rangle + \nonumber \\
&&  \frac{1}{4} N_{\bar{A} \bar{B}} N_{\bar{C} \bar{D}} \int \frac{d^{3} k_3 d^{3} k_4}{(2 \pi)^{6}} \langle \delta \Phi_{\bar{A}} ({\bfk_1} - {\bfk_3}) \delta \Phi_{\bar{B}} ({\bfk_3}) \delta \Phi_{\bar{C}} ({\bfk_2} - {\bfk_4}) \delta \Phi_{\bar{D}} ({\bfk_4}) \rangle - \nonumber \\
&&  \frac{1}{4} N_{\bar{A} \bar{B}} N_{\bar{C} \bar{D}} (2 \pi)^{3} \delta^{3} ({\bfk_2}) \int \frac{d^{3} k_3}{(2 \pi)^{3}} \langle \delta \Phi_{\bar{A}} ({\bfk_1} - {\bfk_3}) \delta \Phi_{\bar{B}} ({\bfk_3}) \rangle \langle \delta \Phi_{\bar{C}} ({\bf x}) \delta \Phi_{\bar{D}} ({\bf x}) \rangle + \nonumber \\
&&  \frac{1}{3!} N_{\bar{A}} N_{\bar{B} \bar{C} \bar{D}} \int \frac{d^{3} k_3 d^{3} k_4}{(2 \pi)^{6}} \langle \delta \Phi_{\bar{A}} ({\bfk_2}) \delta \Phi_{\bar{B}} ({\bfk_1} - {\bfk_3} - {\bfk_4}) \delta \Phi_{\bar{C}} ({\bfk_3}) \delta \Phi_{\bar{D}} ({\bfk_4}) \rangle - \nonumber \\
&&  \frac{1}{4} N_{\bar{A} \bar{B}} N_{\bar{C} \bar{D}} (2 \pi)^{3} \delta^{3} ({\bfk_1}) \int \frac{d^{3} k_3}{(2 \pi)^{3}} \langle \delta \Phi_{\bar{A}} ({\bfk_2} - {\bfk_3}) \delta \Phi_{\bar{B}} ({\bfk_3}) \rangle \langle \delta \Phi_{\bar{C}} ({\bf x}) \delta \Phi_{\bar{D}} ({\bf x}) \rangle + \nonumber \\
&& \frac{1}{4} N_{\bar{A} \bar{B}} N_{\bar{C} \bar{D}} (2 \pi)^{6} \delta^{3} ({\bfk_1}) \delta^{3} ({\bfk_2}) \langle \delta \Phi_{\bar{A}} ({\bf x}) \delta \Phi_{\bar{B}} ({\bf x}) \rangle  \langle \delta \Phi_{\bar{C}} ({\bf x}) \delta \Phi_{\bar{D}} ({\bf x}) \rangle \,.
\end{eqnarray}
The following step is to apply the Wick's theorem \cite{wick}, assuming that the field perturbations may not be gaussian and, again, avoiding terms that would be represented by two- or higher-order loop diagrams:
\begin{eqnarray}
\langle \zeta({\bfk_1}) \zeta({\bfk_2}) \rangle &=& N_{\bar{A}} N_{\bar{B}} \langle \delta \Phi_{\bar{A}} ({\bfk_1}) \delta \Phi_{\bar{B}} ({\bfk_2}) \rangle + \nonumber \\
&&  \frac{1}{2} N_{\bar{A}} N_{\bar{B} \bar{C}} \int \frac{d^{3} k_3}{(2 \pi)^{3}} \langle \delta \Phi_{\bar{A}} ({\bfk_1}) \delta \Phi_{\bar{B}} ({\bfk_2} - {\bfk_3}) \delta \Phi_{\bar{C}} ({\bfk_3}) \rangle_c + \nonumber \\
&&  \frac{1}{3!} N_{\bar{A}} N_{\bar{B} \bar{C} \bar{D}} \int \frac{d^{3} k_3 d^{3} k_4}{(2 \pi)^{6}} \Big[ \langle \delta \Phi_{\bar{A}} ({\bfk_1}) \delta \Phi_{\bar{B}} ({\bfk_2} - {\bfk_3} - {\bfk_4}) \rangle \langle \delta \Phi_{\bar{C}} ({\bfk_3}) \delta \Phi_{\bar{D}} ({\bfk_4}) \rangle + \nonumber \\
&&  \langle \delta \Phi_{\bar{A}} ({\bfk_1}) \delta \Phi_{\bar{C}} ({\bfk_3}) \rangle \langle \delta \Phi_{\bar{B}} ({\bfk_2} - {\bfk_3} - {\bfk_4}) \delta \Phi_{\bar{D}} ({\bfk_4}) \rangle + \nonumber \\
&&  \langle \delta \Phi_{\bar{A}} ({\bfk_1}) \delta \Phi_{\bar{D}} ({\bfk_4}) \rangle \langle \delta \Phi_{\bar{B}} ({\bfk_2} - {\bfk_3} - {\bfk_4}) \delta \Phi_{\bar{C}} ({\bfk_3}) \rangle \Big] + \nonumber \\
&&  \frac{1}{2} N_{\bar{A}} N_{\bar{B} \bar{C}} \int \frac{d^{3} k_3}{(2 \pi)^{3}} \langle \delta \Phi_{\bar{A}} ({\bfk_2}) \delta \Phi_{\bar{B}} ({\bfk_1} - {\bfk_3}) \delta \Phi_{\bar{C}} ({\bfk_3}) \rangle_c + \nonumber \\
&&  \frac{1}{4} N_{\bar{A} \bar{B}} N_{\bar{C} \bar{D}} \int \frac{d^{3} k_3 d^{3} k_4}{(2 \pi)^{6}} \Big[ \langle \delta \Phi_{\bar{A}} ({\bfk_1} - {\bfk_3}) \delta \Phi_{\bar{B}} ({\bfk_3}) \rangle \langle \delta \Phi_{\bar{C}} ({\bfk_2} - {\bfk_4}) \delta \Phi_{\bar{D}} ({\bfk_4}) \rangle + \nonumber \\
&&  \langle \delta \Phi_{\bar{A}} ({\bfk_1} - {\bfk_3}) \delta \Phi_{\bar{C}} ({\bfk_2} - {\bfk_4}) \rangle \langle \delta \Phi_{\bar{B}} ({\bfk_3}) \delta \Phi_{\bar{D}} ({\bfk_4}) \rangle + \nonumber \\
&& \langle \delta \Phi_{\bar{A}} ({\bfk_1} - {\bfk_3}) \delta \Phi_{\bar{D}} ({\bfk_4}) \rangle \langle \delta \Phi_{\bar{B}} ({\bfk_3}) \delta \Phi_{\bar{C}} ({\bfk_2} - {\bfk_4}) \rangle \Big] - \nonumber \\
&&  \frac{1}{4} N_{\bar{A} \bar{B}} N_{\bar{C} \bar{D}} (2 \pi)^{3} \delta^{3} ({\bfk_2}) \langle \delta \Phi_{\bar{C}} ({\bf x}) \delta \Phi_{\bar{D}} ({\bf x}) \rangle \int \frac{d^{3} k_3}{(2 \pi)^{3}} \langle \delta \Phi_{\bar{A}} ({\bfk_1} - {\bfk_3}) \delta \Phi_{\bar{B}} ({\bfk_3}) \rangle  + \nonumber \\
&&  \frac{1}{3!} N_{\bar{A}} N_{\bar{B} \bar{C} \bar{D}} \int \frac{d^{3} k_3 d^{3} k_4}{(2 \pi)^{6}} \Big[ \langle \delta \Phi_{\bar{A}} ({\bfk_2}) \delta \Phi_{\bar{B}} ({\bfk_1} - {\bfk_3} - {\bfk_4}) \rangle \langle \delta \Phi_{\bar{C}} ({\bfk_3}) \delta \Phi_{\bar{D}} ({\bfk_4}) \rangle + \nonumber \\
&&  \langle \delta \Phi_{\bar{A}} ({\bfk_2}) \delta \Phi_{\bar{C}} ({\bfk_3}) \rangle \langle \delta \Phi_{\bar{B}} ({\bfk_1} - {\bfk_3} - {\bfk_4}) \delta \Phi_{\bar{D}} ({\bfk_4}) \rangle + \nonumber \\
&&  \langle \delta \Phi_{\bar{A}} ({\bfk_2}) \delta \Phi_{\bar{D}} ({\bfk_4}) \rangle \langle \delta \Phi_{\bar{B}} ({\bfk_1} - {\bfk_3} - {\bfk_4}) \delta \Phi_{\bar{C}} ({\bfk_3}) \rangle \Big] - \nonumber \\
&&  \frac{1}{4} N_{\bar{A} \bar{B}} N_{\bar{C} \bar{D}} (2 \pi)^{3} \delta^{3} ({\bfk_1}) \langle \delta \Phi_{\bar{C}} ({\bf x}) \delta \Phi_{\bar{D}} ({\bf x}) \rangle \int \frac{d^{3} k_3}{(2 \pi)^{3}} \langle \delta \Phi_{\bar{A}} ({\bfk_2} - {\bfk_3}) \delta \Phi_{\bar{B}} ({\bfk_3}) \rangle + \nonumber \\
&&  \frac{1}{4} N_{\bar{A} \bar{B}} N_{\bar{C} \bar{D}} (2 \pi)^{6} \delta^{3} ({\bfk_1}) \delta^{3} ({\bfk_2}) \langle \delta \Phi_{\bar{A}} ({\bf x}) \delta \Phi_{\bar{B}} ({\bf x}) \rangle  \langle \delta \Phi_{\bar{C}} ({\bf x}) \delta \Phi_{\bar{D}} ({\bf x}) \rangle \,.
\end{eqnarray}
We now employ the definition in Eq. (\ref{ngsp}) for the power spectrum of $\zeta$, and the expressions
\bea\label{psab}
\langle \delta \Phi_{\bar{A}}(\bfk_1)\delta\Phi_{\bar{B}}(\bfk_2)\rangle &\equiv& (2\pi)^3\delta(\bfk_{12})\Pi_{\bar{A}\bar{B}}(\bfk_1) \,, \\ \label{bsabc}
\langle\delta\Phi_{\bar{A}}(\bfk_1)\delta\Phi_{\bar{B}}(\bfk_2)\delta\Phi_{\bar{C}}(\bfk_3)\rangle_c &\equiv& (2\pi)^3\delta(\bfk_{123}) B_{\bar{A}\bar{B}\bar{C}}(\bfk_1,\bfk_2,\bfk_3) \,, \\ \label{tsabcd}
\langle\delta\Phi_{\bar{A}}(\bfk_1)\delta\Phi_{\bar{B}}(\bfk_2)\delta\Phi_{\bar{C}}(\bfk_3)\delta\Phi_{\bar{D}}(\bfk_4)\rangle_c &\equiv& (2\pi)^3\delta(\bfk_{1234})T_{\bar{A}\bar{B}\bar{C}\bar{D}}(\bfk_1,\bfk_2,\bfk_3,\bfk_4) \,,
\eea
and so on; these expressions define the power spectra $\Pi_{\bar{A}\bar{B}}$ of the field perturbations and the higher order functions $B_{\bar{A}\bar{B}\bar{C}}$, $T_{\bar{A}\bar{B}\bar{C}\bar{D}}$, and so on, that parameterize the intrinsic non-gaussianity in the field perturbations.  Thus, calculating some trivial integrals that involve Dirac delta functions we arrive to
\begin{align}
(2 \pi)^{3} \delta^{3} ({\bfk_1} + {\bfk_2}) P_\zeta ({\bfk_1}) =&(2 \pi)^{3} \delta^{3} ({\bfk_1} + {\bfk_2})  \Big[\; N_{\bar{A}} N_{\bar{B}} \Pi_{\bar{A} \bar{B}} ({\bfk_1}) + \frac{1}{2} N_{\bar{A}} N_{\bar{B} \bar{C}} \int \frac{d^{3} k_3}{(2 \pi)^{3}} B_{\bar{A} \bar{B} \bar{C}} ({\bfk_1},{\bfk_2} - {\bfk_3}, {\bfk_3}) + \nonumber \\
&  \frac{1}{2} N_{\bar{A}} N_{\bar{B} \bar{C} \bar{D}} \Pi_{\bar{A} \bar{B}} ({\bfk_1}) \int \frac{d^{3} k_3}{(2 \pi)^{3}} \Pi_{\bar{C} \bar{D}} ({\bfk_3}) + \frac{1}{2} N_{\bar{A}} N_{\bar{B} \bar{C}} \int \frac{d^{3} k_3}{(2 \pi)^{3}} B_{\bar{A} \bar{B} \bar{C}} ({\bfk_2},{\bfk_1} - {\bfk_3}, {\bfk_3}) \; \Big] + \nonumber \\
 &(2 \pi)^{6} \delta^{3} ({\bfk_1}) \delta^{3} ({\bfk_2})  \Big[ \; \frac{1}{4} N_{\bar{A} \bar{B}} N_{\bar{C} \bar{D}} \int \frac{d^{3} k_3 d^{3} k_4}{(2 \pi)^{6}} \Pi_{\bar{A} \bar{B}} ({\bfk_1} - {\bfk_3}) \Pi_{\bar{C} \bar{D}} ({\bfk_2} - {\bfk_4}) \; \Big] + \nonumber \\
 &(2 \pi)^{3} \delta^{3} ({\bfk_1} + {\bfk_2})  \Big[ \; \frac{1}{2} N_{\bar{A} \bar{B}} N_{\bar{C} \bar{D}} \int \frac{d^{3} k_3}{(2 \pi)^{3}} \Pi_{\bar{A} \bar{C}} ({\bfk_1} - {\bfk_3}) \Pi_{\bar{B} \bar{D}} ({\bfk_3}) \; \Big] - \nonumber \\
 &(2 \pi)^{6} \delta^{3} ({\bfk_1}) \delta^{3} ({\bfk_2})  \Big[ \; \frac{1}{4} N_{\bar{A} \bar{B}} N_{\bar{C} \bar{D}} \langle \delta \Phi_{\bar{C}} ({\bf x}) \delta \Phi_{\bar{D}} ({\bf x}) \rangle \int \frac{d^{3} k_3}{(2 \pi)^{3}} \Pi_{\bar{A} \bar{B}} ({\bfk_1} - {\bfk_3}) \; \Big] + \nonumber \\
 &(2 \pi)^{3} \delta^{3} ({\bfk_1} + {\bfk_2})  \Big[ \; \frac{1}{2} N_{\bar{A}} N_{\bar{B} \bar{C} \bar{D}} \Pi_{\bar{A} \bar{B}} ({\bfk_2}) \int \frac{d^{3} k_3}{(2 \pi)^{3}} \Pi_{\bar{C} \bar{D}} ({\bfk_3}) \; \Big] - \nonumber \\
 &(2 \pi)^{6} \delta^{3} ({\bfk_1}) \delta^{3} ({\bfk_2})  \Big[ \; \frac{1}{4} N_{\bar{A} \bar{B}} N_{\bar{C} \bar{D}} \langle \delta \Phi_{\bar{C}} ({\bf x}) \delta \Phi_{\bar{D}} ({\bf x}) \rangle \int \frac{d^{3} k_3}{(2 \pi)^{3}} \Pi_{\bar{A} \bar{B}} ({\bfk_2} - {\bfk_3}) - \nonumber \\
&  \frac{1}{4} N_{\bar{A} \bar{B}} N_{\bar{C} \bar{D}}  \langle \delta \Phi_{\bar{C}} ({\bf x}) \delta \Phi_{\bar{D}} ({\bf x}) \rangle \langle \delta \Phi_{\bar{A}} ({\bf x}) \delta \Phi_{\bar{B}} ({\bf x}) \rangle \; \Big] \,. \nonumber \\
\end{align}

The expression in the third line cancels out exactly the expression in the fifth line; this is because, according to Eq. (\ref{ffe}),
\begin{equation}
\langle \delta \Phi_{\bar{A}} ({\bf x}) \delta \Phi_{\bar{B}} ({\bf x}) \rangle = \int \frac{d^{3} k}{(2 \pi)^{3}} \Pi_{\bar{A} \bar{B}} ({\bf k}) \,.
\end{equation}
The same argument is used to demonstrate that the terms in the last two lines cancel each other out. These four terms are the ones that formally are represented by two-loop disconnected diagrams.  As we see, having defined $\zeta$ as $\delta N - \langle \delta N \rangle$ was key to make this kind of diagrams disappear. Concluding, we may say that the power spectrum of $\zeta$, up to the one-loop level, is given by
\begin{eqnarray}
P_\zeta ({\bfk_1}) &=&
N_{\bar{A}} N_{\bar{B}} \Pi_{\bar{A} \bar{B}} ({\bfk_1}) + \frac{1}{2} N_{\bar{A}} N_{\bar{B} \bar{C}} \int \frac{d^{3} k_3}{(2 \pi)^{3}} B_{\bar{A} \bar{B} \bar{C}} ({\bfk_1},- {\bfk_1} - {\bfk_3}, {\bfk_3}) + \nonumber \\
&&  \frac{1}{2} N_{\bar{A}} N_{\bar{B} \bar{C} \bar{D}} \Pi_{\bar{A} \bar{B}} ({\bfk_1}) \int \frac{d^{3} k_3}{(2 \pi)^{3}} \Pi_{\bar{C} \bar{D}} ({\bfk_3}) + \frac{1}{2} N_{\bar{A}} N_{\bar{B} \bar{C}} \int \frac{d^{3} k_3}{(2 \pi)^{3}} B_{\bar{A} \bar{B} \bar{C}} (-{\bfk_1},{\bfk_1} - {\bfk_3}, {\bfk_3}) + \nonumber \\
&&  \frac{1}{2} N_{\bar{A} \bar{B}} N_{\bar{C} \bar{D}} \int \frac{d^{3} k_3}{(2 \pi)^{3}} \Pi_{\bar{A} \bar{C}} ({\bfk_1} - {\bfk_3}) \Pi_{\bar{B} \bar{D}} ({\bfk_3}) + \frac{1}{2} N_{\bar{A}} N_{\bar{B} \bar{C} \bar{D}} \Pi_{\bar{A} \bar{B}} (-{\bfk_1}) \int \frac{d^{3} k_3}{(2 \pi)^{3}} \Pi_{\bar{C} \bar{D}} ({\bfk_3}) \,. \label{compare1}
\end{eqnarray}

The reader may notice that the obtaining of the previous expression for the power spectrum of $\zeta$ is quite lenghty and clutter, as well as direct but not so intuitive.  As it may be expected, the higher order correlators of $\zeta$ demand even more time and are definitely very clutter leading to undesirable mistakes if extreme care in the manipulation of formulas is not taken; all of this of course means that this methodology of calculation is not in any respect efficient.  The same situation happened in the cannonical formulation of Quantum Field Theory until Feynman presented his diagrammatic rules \cite{feynman1,feynman2,feynman3,diagrammatica}. Following the same idea, the aim of this paper is to present a simple diagrammatic approach to calculate all tree level and loop contributions to the correlation functions 
of $\zeta$ when vector and scalar fields are present. We start in Section \ref{drvs} with the gaussian case and we treat the non-gaussian case in Section \ref{ngdrvs}. We use the tools and the notations that we have introduced in this section throughout the rest of the paper.

\section{Fourier space diagrams for gaussian scalar and vector field perturbations}\label{drvs} 
\begin{figure}[h!] 
   \centering
      \includegraphics[scale=1.5]{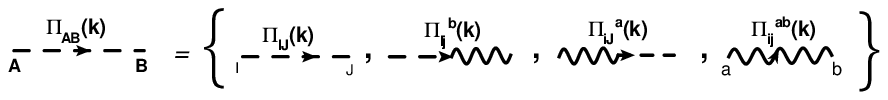}
    \caption{The diagrammatic representation of the propagator: for each pair of indices $\bar{A}, \bar{B}$ we draw a thick dashed line. In order to correctly write the propagator $\Pi_{\bar{A}\bar{B}} ({\bf k})$, the momentum must flow from the end with the label $\bar{A}$ to the end with the label $\bar{B}$; this is relevant if the action is not parity-invariant. In this figure, we show the decomposition of the general propagator in terms of one scalar-scalar propagator, two scalar-vector propagators, and one vector-vector propagator. Thus, for each pair of scalar indices $I, J$ we draw a dashed line, for each pair of scalar-vector indices $I, b$ we draw a dashed-curly line, and for each pair of vector indices $a, b$ we draw a curly line. For each vector index $a$ we understand that there is implicitly an index $i$ corresponding to the components of the vector field. The diagrams in this paper are drawn using JaxoDraw \cite{Binosi:2003yf,Binosi:2008ig}.
}
      \label{prop}
 \end{figure}

\begin{figure}[h!] 
   \centering
      \includegraphics[scale=1.5]{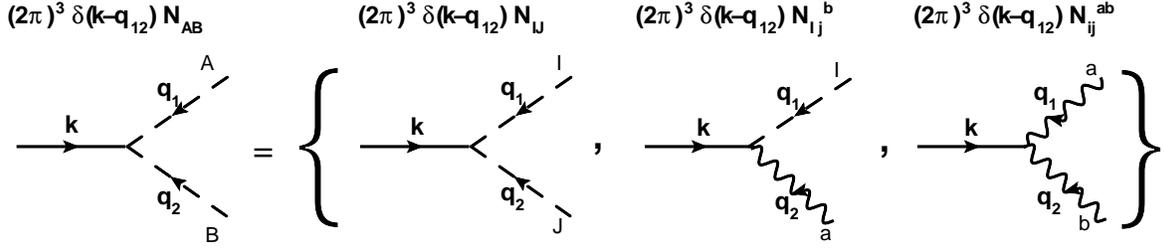}
    \caption{The diagrammatic representation of the 2-vertex $N_{\bar{A}\bar{B}}$. Momentum must be conserved at each vertex: the Dirac delta function guarantees it. Analogously to Fig. \ref{prop}, the 2-vertex has been decomposed  in the scalar-scalar, scalar-vector, and vector-vector 2-vertices.} 
    \label{vert}
 \end{figure}

\begin{figure}[h!]
\centering
\includegraphics[scale=1.5]{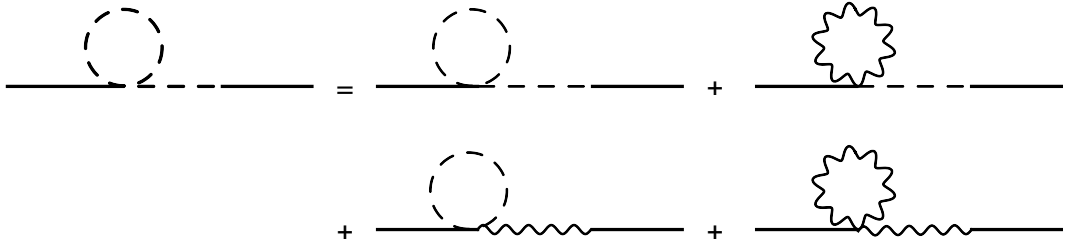}
\caption{A one-loop Feynman-like diagram for power spectrum $P_\zeta$ with gaussian scalar and vector field perturbations (in the decomposition we have assumed isotropic expansion so that there are no scalar-vector propagators). We call it a ``dressed vertex" when the two ends of a propagator are attached to the same vertex as the left one in this figure.}
\label{dressed}
\end{figure}

As we discussed in the previous section, the disconnected  part of the $n$-point correlators of $\zeta$ exactly vanishes due to the substraction of the ensemble average of $\delta N$. We keep this in mind while writing down the rules for drawing the diagrams related to the $n$-point correlator functions of $\zeta$ to the desired order in perturbation theory. 
In this section we present the rules to calculate the $n$-point correlation functions to $p$-order (${\cal O}({P}^{p})$) when $\zeta$ has contributions from scalar and vector field perturbations. Following this terminology, the tree level approximation of the $n$-point correlation function has $p=n-1$ propagators, the 1-loop approximation has $p=n$, and so on.
\begin{enumerate}
\item Draw all distinct diagrams with $n$-external lines and $p$ propagators. 
External lines are represented by solid lines while internal lines are represented by dashed lines.
Every vertex connect an external line to at least 1 propagator.
\item Label the external lines with incoming momenta $\bfk_i$ for $i=1,...,n$ and label the propagator with internal momenta $\bfp_k$ for $k=1,...,p$. 
Label each end of each propagator with a field index: $\bar{A}, \bar{B}, \cdots, \bar{C}$. See Fig. \ref{prop}.
\item Assign the factor $\Pi_{\bar{A}\bar{B}}(\bfk)$ to each propagator in the diagram where $\bfk$ is the momentum associated to the propagator which must flow from the end with the label $\bar{A}$ to the end with the label $\bar {B}$ (see Fig. \ref{prop}). The direction of the momentum flow in the diagram is relevant when the action is not invariant under the parity transformation. If the momentum flows from the end with the label $\bar{B}$ to the end with the label $\bar{A}$, the assigned propagator must be $\Pi_{\bar{B}\bar{A}}(\bfk) = \Pi_{\bar{A}\bar{B}}(-\bfk)$.
\item Assign a factor $N_{\bar{A}_1\bar{A}_2\cdots \bar{A}_t} (2\pi)^3\delta(\bfk_i-\bfp_{1\cdots t})$ to each vertex, a $t$-vertex, where the number $t$ of derivatives of $N$ is the number of propagators attached to this vertex.  We use the convention that incoming momentum is positive. The Dirac delta function ensures that the momentum is conserved. See Fig. \ref{vert}.
\item Integrate over the propagator momenta $\frac{1}{(2\pi)^3}\int d^3p_i$. The first $n-1$ integrals can be done immediately using the Dirac delta functions but any further integral in general cannot be performed analytically. This is the case when there are integrals corresponding to loop corrections.
\item Divide by the appropriate numerical factor:
\begin{itemize}
\item $l!$ if there are $l$ propagators 
attached to the same vertices at both ends. 
\item $2^l l!$ if there are $l$ propagators 
with both ends attached to the same vertex. When the two ends of a propagator attach the same vertex, we say that the propagator is dressing the vertex. See Fig. \ref{dressed}.
\end{itemize}
\item Add all permutations of the diagrams corresponding to all the distinct ways to relabel the $\bfk_i$ attached to the external lines. The number of permutations depends on the symmetries of the diagram: a diagram which is totally symmetric with respect to external lines has only one term; in contrast, a diagram without symmetries with respect to external lines has $n!$ permutations.
\end{enumerate}

In the rules above, we have two basic building blocks. The first one is the Feynman-like propagator $\Pi_{\bar{A}\bar{B}}(\bfk)$:
\be\label{propmultisv}
\Pi_{\bar{A}\bar{B}}(\bfk) \equiv \left\{\Pi_{IJ} (\bfk) \,\, , \,\, \Pi_{Ij}^{\;\; b}(\bfk) \,\, , \,\, \Pi_{iJ}^{a}(\bfk) \,\, , \,\, \Pi_{ij}^{ab}(\bfk) \right\} \,,
\ee
where the $\Pi_{IJ}(\bfk)$ are the scalar-scalar field perturbations power spectra defined by
\begin{equation}
\langle \delta \phi_I (\bfk_1) \delta \phi_J^* (\bfk_2) \rangle = (2 \pi)^3 \delta^3 (\bfk_1 - \bfk_2) \Pi_{IJ} (\bfk_1) \,,
\end{equation}
where we suppose that in general the different scalar fields could be correlated at horizon crossing so that they do not necessarily have the same power spectrum $P_{\delta\phi}$;  they depend on the wavevector since, in general, there will be anisotropic expansion, rendering the field perturbations spectra statistically anisotropic\footnote{The particle production mechanism is statistically anisotropic.}. On the other hand, the $\Pi_{Ij}^{\;\; b}(\bfk)$ are the scalar-vector field perturbations power spectra related to the two-point correlators of a scalar field $\phi_I$ and a vector field $A_j^b$ whose origin lies again in the anisotropic expansion:
\begin{equation}
\langle \delta \phi_I (\bfk_1) \delta A_j^b (\bfk_2) \rangle = (2 \pi)^3 \delta^3 (\bfk_1 + \bfk_2) \Pi_{Ij}^{\; \; b} (\bfk_1) \,.
\end{equation}
Here, $\Pi_{Ij}^{\; \; b} (\bfk)$ is given by:
\be\label{defnew}
\Pi_{Ij}^{\; \; b}(\bfk)\equiv \Pi_{Ij}^{\rm even}(\bfk)P_+^{\; \; b}(\bfk)+i\Pi_{Ij}^{\rm odd}(\bfk)P_-^{\; \; b}(\bfk)+\Pi\su{long}_{Ij}(\bfk)P_{\rm
long}^{\; \; b}(\bfk) \,,
\ee
where it has been written in terms of the longitudinal component of the power spectra $P\sub{long}^{\; \; b}$ and the parity conserving and violating power spectra $P_+^{\; \; b}$ and $P_-^{\; \; b}$ respectively:
\be
P_{\pm}^{\; \; b}\equiv\frac{1}{2}\(P_R^{\; \; b}\pm P_L^{\; \; b}\) \,,
\ee
where $P_R^{\; \; b}$ and $P_L^{\; \; b}$ denote the power spectra for the transverse components with right-handed and left-handed circular polarizations, and again all of them depend on the wavevector because of the anisotropic expansion. The formal definitions of the polarization spectra are given by
\begin{equation}
\langle \delta \phi (\bfk_1) \delta A_\lambda^{b*} (\bfk_2) \rangle = (2 \pi)^3 \delta^3 (\bfk_1 - \bfk_2) P_\lambda^{\; \; b} (\bfk_1) \,,
\end{equation}
where $\lambda$ denotes the different polarizations, $L$, $R$, or $long$, and
\begin{equation}
\delta A_j^b (\bfk) = \sum_\lambda e_j^\lambda (\hat{\bfk}) \delta A_\lambda^b (\bfk) \,,
\end{equation}
${\bf e}^\lambda (\hat{\bfk})$ being the respective polarization vector, and a hat denoting a unit vector.
The basis $\Pi_{Ij}\su{even}$, $\Pi\su{odd}_{Ij}$ and $\Pi\su{long}_{Ij}$ is given by
\be
\Pi_{Ij}\su{even} (\bfk) \equiv \sqrt{2} \hat{p}_j \,,\qquad
\Pi\su{odd}_{Ij} (\bfk) \equiv \sqrt{2} i \epsilon_{jkl} \hat{k}_k \hat{p}_l \,,\qquad
\Pi\su{long}_{Ij} (\bfk) \equiv \hat k_j\,,
\ee
where $\hat{{\bf p}} = \mathcal{R} \hat{{\bf x}}$ and $\hat{{\bf x}}$ being the unit vector that goes in the $x$ direction, $\mathcal{R}$ being the rotation operator that takes $\hat{{\bf z}}$ to $\hat{\bf k}$, and $\hat{{\bf z}}$ being the unit vector that goes in the $z$ direction.
Finally, the $\Pi_{ij}^{ab}(\bfk)$ are the vector-vector field perturbations power spectra defined by:
\begin{equation}
\langle \delta A_i^a (\bfk_1) \delta A_j^b (\bfk_2) \rangle = (2 \pi)^3 \delta^3 (\bfk_1 + \bfk_2) \Pi_{ij}^{a b} (\bfk_1) \,,
\end{equation}
where $\Pi_{ij}^{a b} (\bfk)$ is given by  \cite{dklr}
\be\label{def1}
\Pi_{ij}^{ab}(\bfk)\equiv \Pi_{ij}^{\rm even}(\bfk)P_+^{ab}(\bfk)+i\Pi_{ij}^{\rm odd}(\bfk)P_-^{ab}(\bfk)+\Pi\su{long}_{ij}(\bfk)P_{\rm
long}^{ab}(\bfk) \,,
\ee
$P\sub{long}^{ab}$ being the longitudinal component of the power spectra and $P_+^{ab}$ and $P_-^{ab}$ being the parity conserving and violating power spectra respectively:
\be
P_{\pm}^{ab}\equiv\frac{1}{2}\(P_R^{ab}\pm P_L^{ab}\) \,,
\ee
where $P_R^{ab}$ and $P_L^{ab}$ denote the power spectra for the transverse components with right-handed and left-handed circular polarizations; all of them, as in the other cases, depend on the wavevector because of the anisotropic expansion. Analogously to the previous case, the formal definitions of the polarization spectra are given by
\begin{equation}
\langle \delta A_\lambda^a (\bfk_1) \delta A_\lambda^{b*} (\bfk_2) \rangle = (2 \pi)^3 \delta^3 (\bfk_1 - \bfk_2) P_\lambda^{ab} (\bfk_1) \,.
\end{equation}
The basis $\Pi_{ij}\su{even}$, $\Pi\su{odd}_{ij}$ and $\Pi\su{long}_{ij}$ in this case is given by \cite{dklr}
\be
\Pi_{ij}\su{even} (\bfk) \equiv \delta_{ij} -  \hat k_i \hat k_j \,,\qquad
\Pi\su{odd}_{ij} (\bfk) \equiv \epsilon_{ijk}\hat k_k \,,\qquad
\Pi\su{long}_{ij} (\bfk) \equiv \hat k_i \hat k_j\,.
\ee
We represent diagrammatically the full propagator in Fig. \ref{prop}. 

The second building block of the diagrammatic approach is the $t$-vertex $N_{\bar{A}_1 \cdots \bar{A}_t}$. The diagrammatic representation of the $t$-vertex is shown in Fig. \ref{vert}. There is a factor $(2\pi)^3 \delta(\bfk_i-\bfp_{1\cdots t})$ attached to the vertex. The delta function is needed in order to guarantee momentum conservation at the vertex.  $\bfk_i$ is an external momentum (bold solid line), and the $\bfp_k$ are the internal momenta (bold dashed line).

Examples of the diagrams derived from these rules are discussed in the next section.





\subsection{Examples}
In this section we show how the diagrammatic rules work in a particular case. We will assume that the  curvature perturbation is generated by a single scalar and a single vector field so that $\delta\Phi_{\bar{A}}= (\delta\phi\, ,\, \delta A_i)$ and the $\dn$ series goes as follows:
\be
\zeta(\bfx , t)\equiv\delta N (\phi(\bfx),A_i(\bfx),t)=N_\phi \delta\phi + N_i\delta A_i+\frac{1}{2}N_{\phi\phi}
(\delta\phi)^2+
N_{\phi i}\delta\phi\delta A_i+\frac{1}{2}N_{ij}\delta A_i \delta A_j \,+ \cdots \label{dNsv} - ({\rm spatial \ average}) \,.
\ee
According to Eq. (\ref{dNsv}) the $n$-point correlation functions have contributions from the vector field perturbations, from the scalar field perturbations and from  mixed terms. We will show how the rules work in this case for the two-point, the three-point, and the four-point correlators of $\zeta$. We will work out in detail the case of the 1-loop power spectrum and describe schematically the other cases.
\subsubsection{The tree level and one-loop contributions to the spectrum $\pz$}   
For the power spectrum we have $n=2$, hence, the tree level approximation has $p=2-1=1$ propagator. At this level we then have only two first order derivatives (1-vertex) and one propagator joining them. 
Diagrammatically the two-point correlator of $\zeta$ is represented  by:
\be
\begin{array}{lll}
\langle\zeta({\bfk_1})\zeta({\bfk_2})\rangle^{\rm tree} & = & \includegraphics[scale=0.91]{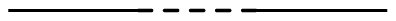} \\
 &  = & \\
 &  &  \includegraphics[scale=0.95]{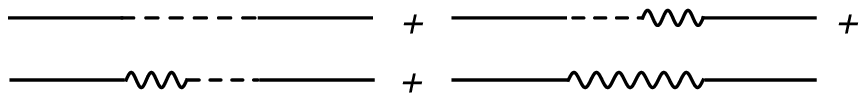} \,.
\end{array}
\ee	
Using the rules given at the beginning of this section, and after doing the trivial integration, the tree level term to the power spectrum $P_\zeta$ is given by:
\be\label{ptree}
\pz^{\rm tree}({\bfk_1}) = N_{\bar{A}}N_{\bar{B}}\Pi_{\bar{A}\bar{B}}({\bfk_1}) = N_{\phi}^2 P(\bfk_1) + N_{\phi} N_j \Pi_{\phi j} (\bfk_1) + N_{\phi} N_j \Pi_{\phi j} (-\bfk_1) + N_iN_j\Pi_{ij}({\bfk_1}) \,,
\ee
where $P(\bfk)$ is the scalar-scalar field power spectrum, $\Pi_{\phi j} (\bfk)$ is the scalar-vector field propagator, and $\Pi_{ij}(\bfk)$ is the vector-vector field propagator.

Going now to the one-loop correction, we have $n=2$ external lines  and $p=2$ propagators. We have one term with two 2-vertices attached to the two external lines and the two propagators  joining the vertices. We also have another term with one 1-vertex and one 3-vertex, one propagator joins the vertices and the other one dresses the 3-vertex. 
Assuming isotropic expansion, the one-loop contribution may easily be diagrammatically expanded since we avoid scalar-vector propagators; in this case the resulting diagrammatic representation is:
\bea
\langle\zeta({\bfk_1})\zeta({\bfk_2})\rangle^{\rm 1-loop}\ &=& \ \begin{array}{l}
\includegraphics[scale=0.85]{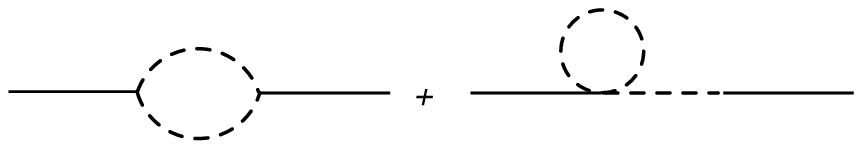} \nonumber
\end{array}\\
\ &=& \ \begin{array}{l}
\includegraphics[scale=0.85]{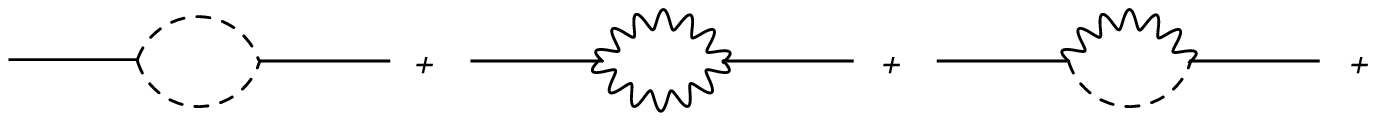} \nonumber
\end{array}\\ 
\ && \ \begin{array}{l}
\includegraphics[scale=0.85]{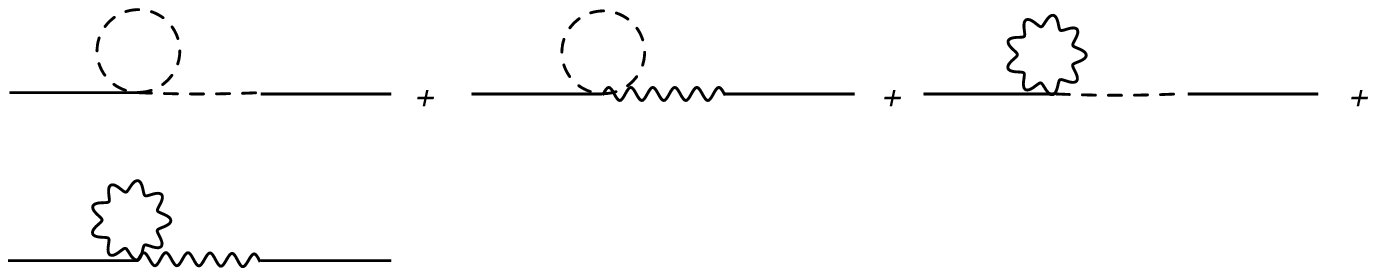} \,.
\end{array}
\eea
Here we can identify seven different diagrams associated to contributions from scalar, vector and mixed terms. Again, using the rules given at the beginning of this section, it is easy to get the one-loop correction to $P_\zeta$. After carrying out one integration (the trivial one) we get:
\bea
P_\zeta^{\rm 1-loop} ({\bfk_1}) &=&  \frac{1}{2} N_{\bar{A}\bar{B}}N_{\bar{C}\bar{D}} \int \frac{d^3p}{(2\pi)^3}  \Pi_{\bar{A}\bar{C}}({\bfk_1}+\bfp)\Pi_{\bar{B}\bar{D}}(-\bfp) + \nonumber \\
&&   \frac{1}{2} N_{\bar{A}}N_{\bar{B}\bar{C}\bar{D}} \int \frac{d^3p}{(2\pi)^3}  \Pi_{\bar{A}\bar{B}}(-{\bfk_1})\Pi_{\bar{C}\bar{D}}(\bfp)  +  \frac{1}{2} N_{\bar{A}}N_{\bar{B}\bar{C}\bar{D}} \int \frac{d^3p}{(2\pi)^3}  \Pi_{\bar{A}\bar{B}}({\bfk_1})\Pi_{\bar{C}\bar{D}}(\bfp) \label{compare} \\
&=& \int \frac{d^3p}{(2\pi)^3} \[\frac{1}{2} N_{\phi \phi}^2  P (|{\bfk_1} + \bfp|)
P(p) + \frac{1}{2} N_{ij} N_{kl}\Pi_{ik}({\bfk_1}+\bfp)\Pi_{jl}(-\bfp)  +N_{\phi i} N_{\phi j} P (|{\bfk_1} + \bfp|)
\Pi_{ij}(-\bfp)  \] + \nonumber \\
&& \int \frac{d^3p}{(2\pi)^3} \Big[ N_{\phi } N_{\phi \phi \phi}  P (k_1)P (p) + \frac{1}{2} N_{i } N_{j \phi \phi }  P (p)\Pi_{ij}(-{\bfk_1}) + \frac{1}{2} N_{i } N_{j \phi \phi }  P (p)\Pi_{ij}({\bfk_1}) +  \nonumber \\
&& N_{\phi } N_{\phi ij}  P (k_1)\Pi_{ij} (\bfp) + \frac{1}{2} N_{i } N_{jkl}  \Pi_{ij}(-{\bfk_1})\Pi_{kl}(\bfp) + \frac{1}{2} N_{i } N_{jkl}  \Pi_{ij}({\bfk_1})\Pi_{kl}(\bfp)  \Big] \,.
\label{Pzetal}
\eea
In the first two lines we have written the total one-loop contribution in terms of the t-vertices $N_{\bar{A}_1\cdots \bar{A}_t}$ and the propagators $\Pi_{\bar{A}\bar{B}}$ and in the lines below we have expanded them in terms of their scalar-scalar, scalar-vector, and vector-vector components. Notice that the first term has a numerical factor $1/2$ because, according to rule 6, there are two propagators attached to the same vertex. The second and third terms also have a factor $1/2$ because there is a propagator dressing a vertex.  When we do the expansion of the first term in the scalar and vector components the terms with pure-scalar fields and pure-vector fields will retain the numerical factor $1/2$, however the term associated to mixed contributions (scalar and vector) does not have this numerical factor; this is because the propagators are not of the same kind. Similar considerations apply for those terms in the last two lines. 

Comparing Eqs. (\ref{compare1}), (\ref{ptree}) and (\ref{compare}) we observe that the diagrammatic approach is able to reproduce the power spectrum of $\zeta$ up to the one-loop approximation and assuming that the field perturbations obey a gaussian statistics.  Nevertheless, the application of the diagrammatic method is very quick, useful, and intuitive, in absolut contrast with the traditional method employing the Wick's theorem.  Here is where the power of the Feynman-like rules relies.
\subsubsection{The tree level and one-loop contributions to the bispectrum $\bz$}
\begin{figure}[h!]
\centering
\includegraphics[scale=1]{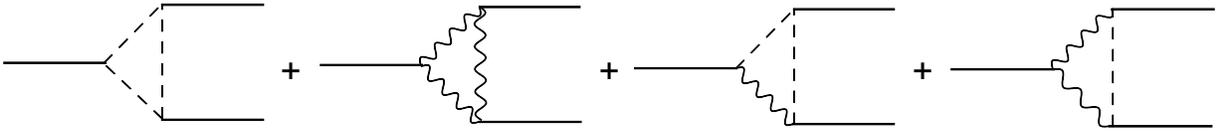}
\caption{The scalar and vector components in the expansion of the $\bz^{222}$ term assuming isotropic expansion.}
\label{b222}
\end{figure}

Assuming $\zeta$ as in \eq{dNsv} and using the rules  given at the beginning of this section to draw the diagrams, the tree level term $(p=3-1=2)$ to the three-point correlation function of $\zeta$ is diagrammatically given by:
\bea\nonumber
\langle\zeta(\bfk_1)\zeta(\bfk_2)\zeta(\bfk_3)\rangle^{\rm tree}\ &=& \ \begin{array}{l}
\includegraphics[scale=0.8]{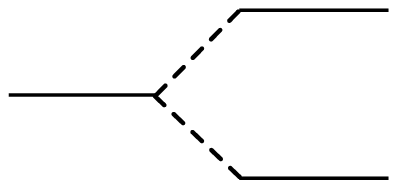}
\end{array}\\
\ &=& \ \begin{array}{l}
\includegraphics[scale=0.8]{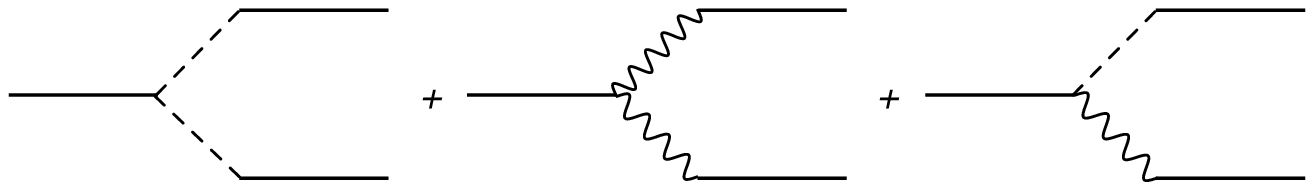} \,,
\end{array}
\eea
where the diagrammatic expansion has been done assuming isotropic expansion, i.e., avoiding scalar-vector propagators.
This diagram is the result of joining three external lines with two 1-vertices and one 2-vertex with $p=2$ propagators. 
Using the diagrammatic rules we obtain the expression for the tree level bispectrum:
\bea
\bz^{\rm tree} (\bfk_1,\bfk_2,\bfk_3) &=& N_{\bar{A}} N_{\bar{B}} N_{\bar{C}\bar{D}}
\Big[ \Pi_{\bar{A}\bar{C}}(\bfk_1)\Pi_{\bar{B}\bar{D}}(\bfk_2) + 2 \ {\rm perm.}\Big]  \no\\
 &=& N_{\phi}^2 N_{\phi \phi}
[P(k_1) P (k_2) + 2 \ {\rm perm.}]  + N_i N_k N_{mn}
\Big[ \Pi_{im}(\bfk_1)\Pi_{kn}(\bfk_2) + 2 \ {\rm perm.}\Big] + \no\\
 && N_{\phi} N_i N_{\phi j} \Big[P(k_1)
\Pi_{ij}(\bfk_2) + 5 \ {\rm perm.} \Big] \,. \label{bzt}
\eea \\
The one-loop contribution $p=3$ is represented by
\bea\nonumber
\langle\zeta(\bfk_1)\zeta(\bfk_2)\zeta(\bfk_3)\rangle^{\rm 1-loop}\ &=& \ \begin{array}{l}
\includegraphics[scale=0.9]{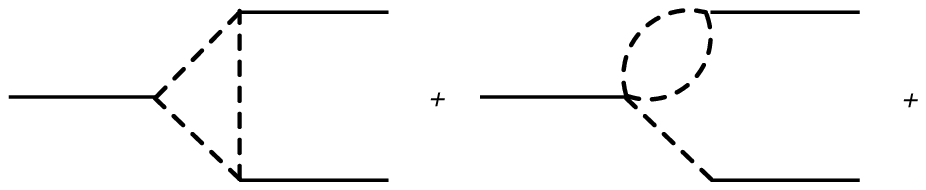}
\end{array}\\
\ && \ \begin{array}{l}
\includegraphics[scale=0.9]{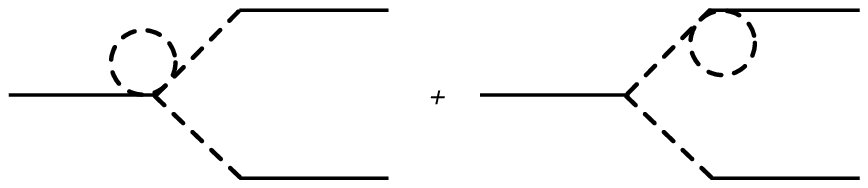} \,,
\end{array}
\eea
which is the result of joining three propagators with three external lines.  There are four different ways of joining the propagators and the external lines: we have the first term with three 2-vertices, the second term which contains one 1-vertex, one 2-vertex and one 3-vertex, the third term with two 1-vertices and one dressed 4-vertex, and the fourth term with one 1-vertex, one 2-vertex and one dressed 3-vertex.  Again, using the rules and performing the trivial integrations (thanks to the Dirac delta function), the above diagrams lead to the following expression:
\bea \label{bs-zeta} \nonumber
\bz^{\rm 1-loop}(\bfk_1,\bfk_2,\bfk_3)&=&   N_{\bar{A}\bar{B}} N_{\bar{C}\bar{D}} N_{\bar{E}\bar{F}} \int \frac{d^3p}{(2\pi)^3} \;\; \Pi_{\bar{A}\bar{C}}(-\bfp) \Pi_{\bar{B}\bar{E}}(\bfk_1 + \bfp) \Pi_{\bar{D}\bar{F}}(\bfk_2 - \bfp) +
\\ \nonumber
&&  \frac{1}{2}N_{\bar{A}} N_{\bar{B}\bar{C}} N_{\bar{D}\bar{E}\bar{F}}\left[ \int \frac{d^3p}{(2\pi)^3} \;\; \Pi_{\bar{A}\bar{D}}(\bfk_1) \Pi_{\bar{B}\bar{E}}(\bfp) \Pi_{\bar{C}\bar{F}}(\bfk_2 - \bfp) + {\mbox 5\, {\rm perm.}} \right]  + \\ \nonumber
&&  \frac{1}{2}N_{\bar{A}} N_{\bar{B}} N_{\bar{C}\bar{D}\bar{E}\bar{F}} \left[\int \frac{d^3p}{(2\pi)^3} \;\;\Pi_{\bar{A}\bar{C}}(\bfk_1) \Pi_{\bar{B}\bar{D}}(\bfk_{2} )  \Pi_{\bar{E}\bar{F}}(\bfp) + {\mbox 2\, {\rm perm.}} \right] + \\
&& \frac{1}{2}N_{\bar{A}} N_{\bar{B}\bar{C}} N_{\bar{D}\bar{E}\bar{F}}\left[ \int \frac{d^3p}{(2\pi)^3}\;\;\Pi_{\bar{A}\bar{B}}(\bfk_1) \Pi_{\bar{C}\bar{D}}(\bfk_{12} )  \Pi_{\bar{E}\bar{F}}(\bfp) + {\mbox 5\, {\rm perm.}} \right].
\eea
We have to expand every term in its scalar-scalar, scalar-vector, and vector-vector components. 
As an example, we expand the first term which has three 2-vertices in the isotropic expansion limit. We will call this term $\bz^{222}$ because it has three 2-vertices:
\bea
\bz^{222}(\bfk_1,\bfk_2,\bfk_3) &=& N_{\bar{A}\bar{B}} N_{\bar{C}\bar{D}} N_{\bar{E}\bar{F}} \int \frac{d^3p}{(2\pi)^3} \;\; \Pi_{\bar{A}\bar{C}}(-\bfp) \Pi_{\bar{B}\bar{E}}(\bfk_1 + \bfp) \Pi_{\bar{D}\bar{F}}(\bfk_2 - \bfp) \nonumber
\\ \nonumber
&=& N_{\phi \phi}^3\int\frac{d^3p}{(2\pi)^3}P(p)P(|\bfk_1+\bfp|)
P(|\bfk_2-\bfp|) + \no\\
&& N_{ij}N_{kl}N_{mn}\int\frac{d^3p}{(2\pi)^3}\Pi_{ik}(-\bfp)\Pi_{jm}
(\bfk_1+\bfp)\Pi_{ln}(\bfk_2-\bfp) + \no\\
&&N_{\phi \phi}N_{\phi i}N_{\phi j}\int\frac{d^3p}{(2\pi)^3}\bigg\{P(p)P(|\bfk_1+\bfp|)\Pi_{ij}(\bfk_2-\bfp) + \no\\
&&P(p)P(|\bfk_2-\bfp|)\Pi_{ij}(\bfk_1+\bfp)+P(|\bfk_2-\bfp|)P(|\bfk_1+\bfp|)\Pi_{ij}(\bfp)\bigg\}+\no\\
&&N_{\phi i}N_{\phi j}N_{kl}\int\frac{d^3p}{(2\pi)^3}\bigg\{P(p)\Pi_{ik}(\bfk_1+\bfp)\Pi_{jl} (\bfk_2-\bfp) + \no\\
&&P(|\bfk_1+\bfp|)\Pi_{ik}(\bfp)\Pi_{jl}(\bfk_2-\bfp)+P(|
\bfk_2-\bfp|)\Pi_{ik}(\bfp)\Pi_{jl}(\bfk_1+\bfp)\bigg\}\label{bsl} \,.
\eea
The diagrammatic representation of this expansion is shown in Fig. \ref{b222}.
\subsubsection{The tree level contribution to the trispectrum $\tz$}  
Finally, we draw the diagrams and write down the expression for the tree level term of the trispectrum $\tz$. This term has $p=4-1=3$ propagators joining $n=4$ external legs. There are two ways of joining the three propagators with four external lines as we see in its diagrammatic representation:
\bea
\langle\zeta(\bfk_1)\zeta(\bfk_2)\zeta(\bfk_3)\zeta(\bfk_4)\rangle^{\rm tree} \ &=& \ \begin{array}{l}
\includegraphics[scale=1]{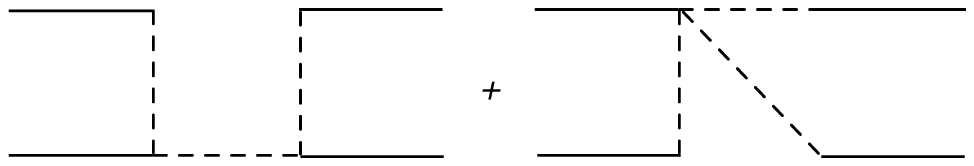} \,.
\end{array}
\eea
Using the rules in Section \ref{drvs} we obtain the tree level expression for the trispectrum:
\bea \label{tst}
\tz^{\rm tree} (\bfk_1,\bfk_2,\bfk_3,\bfk_4) &=& N_{\bar{A}} N_{\bar{B}} N_{\bar{C}\bar{D}} N_{\bar{E}\bar{F}} \left[\Pi_{\bar{A}\bar{C}}(\bfk_1) \Pi_{\bar{B}\bar{E}}(\bfk_2)\Pi_{\bar{D}\bar{F}}(\bfk_{13}) + {\mbox 11\, {\rm perm.}} \right] + \nonumber \\
&&  N_{\bar{A}} N_{\bar{B}} N_{\bar{C}} N_{\bar{D}\bar{E}\bar{F}}  \left[\Pi_{\bar{A}\bar{D}}(\bfk_1) \Pi_{\bar{B}\bar{E}}(\bfk_2)\Pi_{\bar{C}\bar{F}}(\bfk_3) + {\mbox 3\, {\rm perm.}} \right].
\eea
\section{Diagrammatic rules for non-gaussian vector and scalar field perturbations}\label{ngdrvs} 
\begin{figure}[h] 
   \centering
      \includegraphics[scale=1.2]{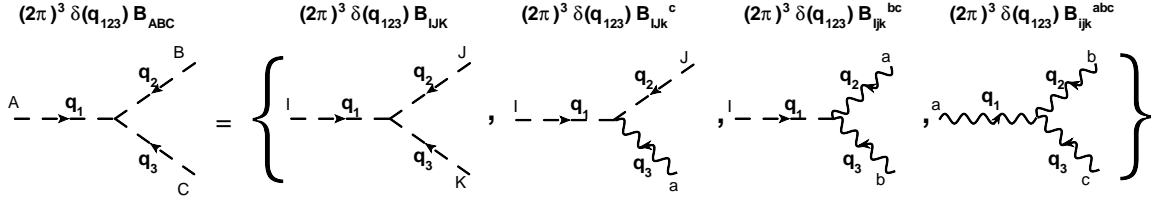}
    \caption{The diagrammatic representation of the bispectrum in the field perturbations. Incoming momenta are considered to be positive.  In the diagrammatic decomposition, for each scalar index $I$ we draw a dashed line while for each vector index $a$ we draw a curly line. For each vector index $a$ we understand that there is implicitly an index $i$ corresponding to the components of the vector field.    \label{grafbsabc}}
 \end{figure}

In this section we extend the rules obtained in Section \ref{drvs} in order to include the contributions coming from non-gaussian correlators of the field perturbations. We include the non-gaussianity in the form of primordial connected $k$-point correlation functions which we will call $(k-1)$-spectra.  With this notation $k=2$ is a propagator. As we have studied in Section \ref{drvs}, in the gaussian case we have only propagators and $t$-vertices in the diagrammatic perturbative series.  As an example of primordial non-gaussian contributions we have the terms containing the 3-point correlators of the field perturbations, the 2-spectrum (or bispectrum) $B_{\bar{A}\bar{B}\bar{C}}$ in Eq. (\ref{bsabc}). In this case, aside of the propagator and the $t$-vertex we have to add the bispectrum of the field perturbations to the building block list. The diagrammatic representation of the bispectrum in the field perturbations is shown in Fig. \ref{grafbsabc}.
The rules to evaluate perturbatively a $n$-point correlation function of $\zeta$, including non-gaussianity in the field perturbations, are constructed by adding the $(k-1)$-spectra to the building block list with the appropriate rules to deal with their symmetries in a diagram. They are listed as follows:
\begin{enumerate}
\item Draw all distinct diagrams with $n$-external lines, and the appropriate number $s$ of $(k-1)$-spectra. The case $k=2$ corresponds to a propagator. The order of a $(k-1)$-spectrum in the perturbative expansion is $k-1$, this is $\langle \delta \Phi_{\bar{A}_1}\cdots\delta\Phi_{\bar{A}_k}\rangle\sim {\cal O}({P}^{k-1})$.  
External lines are represented by solid lines while internal lines are represented by dashed lines. 
 Every vertex cannot have attached more than one external line. Every line in a $(k-1)$-spectra must be attached on at least one end to an external line, if it is attached to external lines in both ends it is a propagator.
\item Label the external lines with incoming momenta $\bfk_i$ for $i=1,...,n$ and label the $(k-1)$-spectra with internal momenta $\bfp_k$ for $k=1,...,p$. Label each end of each $(k-1)$-spectra with a field index: $\bar{A}, \bar{B}, \cdots, \bar{C}$. 
See Figs. \ref{prop} and \ref{grafbsabc}.
\item Assign the factor $\Pi_{\bar{A}\bar{B}}(\bfk)$ to each propagator in the diagram where $\bfk$ is the momentum associated to the propagator which must flow from the end with the label $\bar{A}$ to the end with the label $\bar {B}$ (see Fig. \ref{prop}). The direction of the momentum flow in the diagram is relevant when the action is not invariant under the parity transformation. If the momentum flows from the end with the label $\bar{B}$ to the end with the label $\bar{A}$, the assigned propagator must be $\Pi_{\bar{B}\bar{A}}(\bfk) = \Pi_{\bar{A}\bar{B}}(-\bfk)$.
\item Assign the factor $F_{\bar{A}_1\bar{A}_2\cdots\bar{A}_k}(\bfk_1 , \bfk_2,\cdots , \bfk_k)(2\pi)^3\delta(\bfk_{12\cdots k})$ to each primordial $(k-1)$-spectrum in the diagram. $\bfk_1 , \bfk_2, \cdots \bfk_k$ are the incoming momenta associated to the $(k-1)$-spectrum.  See for example Fig. \ref{grafbsabc} for the case of the bispectrum. The Dirac delta function ensures that the momentum is conserved.
\item Assign a factor $N_{\bar{A}_1\bar{A}_2\cdots \bar{A}_t} (2\pi)^3\delta(\bfk_i-\bfp_{1\cdots t})$ to each vertex, a $t$-vertex, where the number $t$ of derivatives of $N$ is the number of 
lines attached to this vertex.  We use the convention that incoming momentum is positive. The Dirac delta function ensures that the momentum is conserved. See Fig. \ref{vert}.
\item Integrate over the propagator momenta $\frac{1}{(2\pi)^3}\int d^3p_i$. The first $n-1$ integrals can be done immediately using the Dirac delta functions but any further integral in general cannot be performed analytically. This is the case when there are integrals corresponding to loop corrections.
\item Divide by the appropriate numerical factor:
\begin{itemize}
\item $l!$ if there are $l$ propagators 
attached to the same vertices at both ends.
\item $2^l l!$ if there are $l$ propagators 
dressing a vertex.
\item $p!$ if there are $p$ $(k-1)$-spectra 
attached to the same vertices at both ends, and with just one line attached to each vertex.
\item $m!n!...$ if there are $m$ lines of just one $(k-1)$-spectrum  attached to the same vertex $V_1$, and $n$ lines of the same $(k-1)$-spectrum attached to the same vertex $V_2$, and so on.
\item $p! (m!n!...)^p$ if there are $m$ lines of each of the $p$ $(k-1)$-spectra 
 attached to the same vertex $V_1$, and $n$ lines of each of the same $p$ $(k-1)$-spectra attached to the same vertex $V_2$, and so on.
\item $l!((k-1)!)^{l}$ if there are $l$ $(k-1)$-spectra all of the same order 
dressing a vertex.
\end{itemize}
\item Add all permutations of the diagrams corresponding to all the distinct ways to relabel the $\bfk_i$ attached to the external lines. The number of permutations depends on the symmetries of the diagram: a diagram which is totally symmetric with respect to external lines has only one term; in contrast, a diagram without symmetries with respect to external lines has $n!$ permutations.
\end{enumerate}

In this case, given that the $(k-1)$-spectra are part of the building blocks of the diagrammatic representation, it is instructive to know the maximum number of them that we can have in every $l$-loop term. This information is relevant in order to construct all the possible diagrams at each order. In the process of drawing a $l$-loop diagram we have to take into account that the sum of the order of all the $(k-1)$-spectra in a diagram must coincide with the order of the diagram.  In Table 1 we summarize the maximum number of $(k-1)$-spectra that could appear in any diagram corresponding to a $l$-loop correction in the perturbative series for a $n$-point correlation function.
\begin{table}[h]\label{table_loop}
  \caption{The number of $(k-1)$-spectra that we need to construct the $l$-loop term of a $n$-point correlation function. In the rows we have the $l$-loop order correction and in the columns we show the maximum number of $(k-1)$-spectra that could appear at each order. $ \[x\] $ represents the integer part of $x$.  }
\begin{center}{\small
\begin{tabular}{ | c | c | c | c | c | c |}
    \hline
    loop order & spectrum & 2-spectrum   & 3-spectrum & $\cdots$ & $(k-1)$-spectrum   \\ \hline \hline
    tree: $n-1$ & $n-1$ & $\[(n-1)/2\]$   & $\[(n-1)/3\]$ & $\cdots$ & $\[(n-1)/(k-1)\]$    \\ \hline
    1-loop: $n$ & $n$ & $\[n/2\]$   & $\[n/3\]$ & $\cdots$ & $\[n/(k-1)\]$    \\ \hline
    2-loops: $n+1$ & $n+1$ & $\[(n+1)/2\]$   & $\[(n+1)/3\]$ & $\cdots$ & $\[(n+1)/(k-1)\]$    \\ \hline
    $\cdots$ & $\cdots$ & $\cdots$   & $\cdots$ & $\cdots$ & $\cdots$   \\ \hline
    $l$-loops: $n+l-1$ & $n+l-1$ & $\[(n+l-1)/2\]$   & $\left[(n+l-1)/3\right]$ & $\cdots$ & $ \[ (n+l-1)/(k-1) \right]$    \\ \hline
    \end{tabular}}
    \end{center}
\end{table}
\subsection{Examples}
Again, we are going to use the example of a single scalar field and a single vector field studied in Section \ref{drvs}. We will work out in detail the power spectrum and describe briefly the bispectrum and trispectrum.

\subsubsection{The tree level and one-loop contributions to the spectrum $\pz$}   
For the power spectrum we have $n=2$ external lines, and the tree level approximation has $p=2-1=1$ propagator, {\it i.e} it is of order ${\cal O} ({P}^1)$. Higher order primordial $(k-1)$-spectra do not appear at this order because the lowest higher-order correlator is the bispectrum which is of order ${\cal O} ({P}^2)$. Hence, the tree level power spectrum is identical to the one in Eq. (\ref{ptree}). The 1-loop power spectrum is of order  ${\cal O} ({P}^2)$ so, according to Table 1, we have two propagators and one bispectrum. We do not have higher-order correlators of the field perturbations at this order. The diagrammatic representation of the 1-loop power spectrum is:
\bea
\langle\zeta({\bfk_1})\zeta({\bfk_2})\rangle^{\rm 1-loop}\ &=& \ \begin{array}{l}
\includegraphics[scale=0.8]{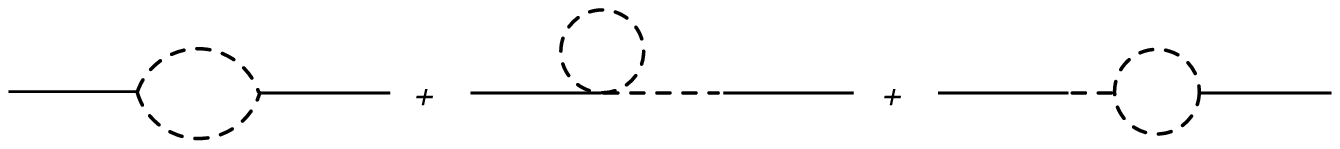} \,.
\end{array}
\eea
Using the diagrammatic rules we find the expression for the 1-loop power spectrum:
\bea
P_\zeta^{\rm 1-loop} ({\bfk_1}) &=&  \frac{1}{2} N_{\bar{A}\bar{B}}N_{\bar{C}\bar{D}} \int \frac{d^3p}{(2\pi)^3}  \Pi_{\bar{A}\bar{C}}({\bfk_1}+\bfp)\Pi_{\bar{B}\bar{D}}(-\bfp) + \nonumber \\
&&  \frac{1}{2}N_{\bar{A}}N_{\bar{B}\bar{C}\bar{D}} \int \frac{d^3p}{(2\pi)^3}  \Pi_{\bar{A}\bar{B}}(-{\bfk_1})\Pi_{\bar{C}\bar{D}}(\bfp) + \frac{1}{2}N_{\bar{A}}N_{\bar{B}\bar{C}\bar{D}} \int \frac{d^3p}{(2\pi)^3}  \Pi_{\bar{A}\bar{B}}({\bfk_1})\Pi_{\bar{C}\bar{D}}(\bfp) + \no\\
&& \frac{1}{2} N_{\bar{A}}N_{\bar{B}\bar{C}} \int \frac{d^3p}{(2\pi)^3}  B_{\bar{A}\bar{B}\bar{C}}({\bfk_1}, \bfp, -{\bfk_1} - \bfp) + \frac{1}{2} N_{\bar{A}}N_{\bar{B}\bar{C}} \int \frac{d^3p}{(2\pi)^3}  B_{\bar{A}\bar{B}\bar{C}}(-{\bfk_1}, \bfp, {\bfk_1} - \bfp) \,. \nonumber \\
\eea
Comparing with the gaussian case in Eq. (\ref{compare}), the 1-loop power spectrum has the terms in the last line as extra terms.  These terms have a bispectrum of the field perturbations connected with a 1-vertex and a 2-vertex.
Comparing again this expression with Eq. (\ref{compare1}), we see that the one-loop contribution to the power spectrum of $\zeta$ calculated using the usual approach, employing the Wick's theorem, is fully reproduced, but this time with much less effort.

\subsubsection{The tree level and one-loop contributions to the bispectrum $\bz$}
According to Table 1, for the tree level 3-point correlation function we have a maximum of 2 propagators and one bispectrum of the field perturbations. With them, we can construct the following diagrams:
\bea
\langle\zeta(\bfk_1)\zeta(\bfk_2)\zeta(\bfk_3)\rangle^{\rm tree}\ &=& \ \begin{array}{l}
\includegraphics[scale=0.8]{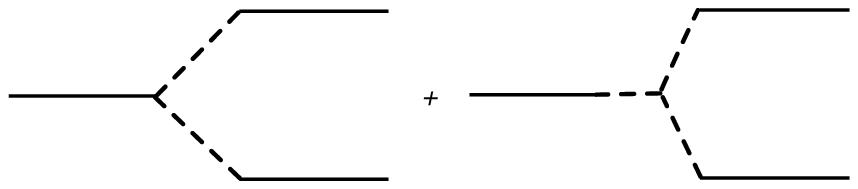}
\end{array}.
\eea
Comparing with the gaussian expression, we have a new term coming from the bispectrum of the field perturbations. Using the diagrammatic rules we obtain the expression for the tree level bispectrum:
\bea
\bz^{\rm tree} (\bfk_1,\bfk_2,\bfk_3) &=& N_{\bar{A}} N_{\bar{B}} N_{\bar{C}\bar{D}}
\Big[ \Pi_{\bar{A}\bar{C}}(\bfk_1)\Pi_{\bar{B}\bar{D}}(\bfk_2) + 2 \ {\rm perm.}\Big]   +  N_{\bar{A}} N_{\bar{B}} N_{\bar{C}} B_{\bar{A}\bar{B}\bar{C}}(\bfk_1,\bfk_2,\bfk_3)\,. \label{bztng}
\eea \\
The one-loop contribution is constructed out of the possible diagrams that we can draw with a maximum of $p=3$ propagators, $b=2$ bispectra and $t=1$ trispectrum. The resulting diagrams are shown here:
\bea
\langle\zeta(\bfk_1)\zeta(\bfk_2)\zeta(\bfk_3)\rangle^{\rm 1-loop}\ &=& \ \begin{array}{l}
\includegraphics[scale=0.68]{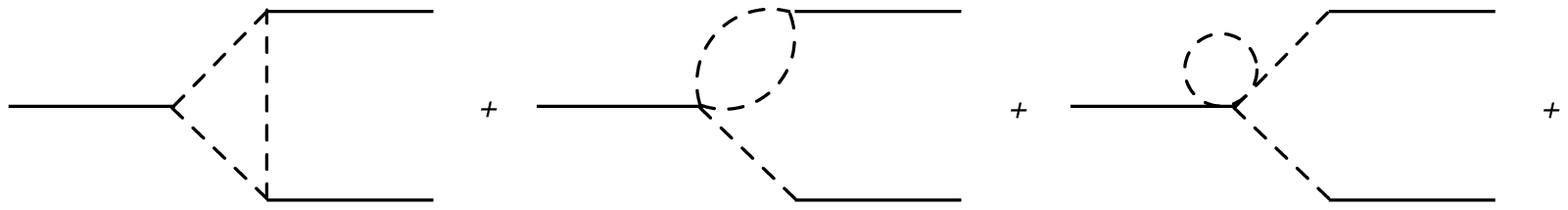}
\end{array}\\ \nonumber
\ && \ \begin{array}{l}
\includegraphics[scale=0.78]{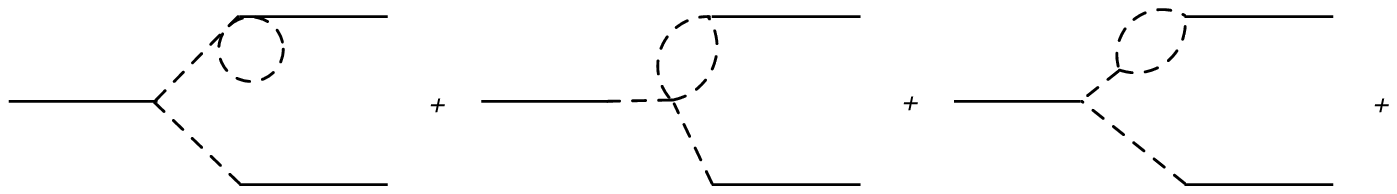}
\end{array}\\ \nonumber
&& \ \begin{array}{l}
\includegraphics[scale=0.78]{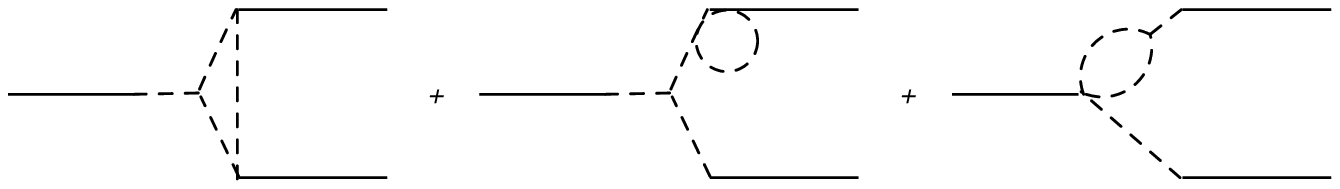} \,.
\end{array} \
\eea
We can see that, compared with the gaussian case, there are five new terms related to the bispectrum and trispectrum of the field perturbations. Again, using the rules and doing the trivial Dirac delta function integrations, the above diagrams lead to the following expressions:
\bea \label{bs-zetang}
\bz^{\rm 1-loop}(\bfk_1,\bfk_2,\bfk_3)&=&   N_{\bar{A}\bar{B}} N_{\bar{C}\bar{D}} N_{\bar{E}\bar{F}} \int \frac{d^3p}{(2\pi)^3} \;\; \Pi_{\bar{A}\bar{C}}(-\bfp) \Pi_{\bar{B}\bar{E}}(\bfk_1 + \bfp) \Pi_{\bar{D}\bar{F}}(\bfk_2 - \bfp) + \nonumber \\
&&  \frac{1}{2}N_{\bar{A}} N_{\bar{B}\bar{C}} N_{\bar{D}\bar{E}\bar{F}}\left[ \int \frac{d^3p}{(2\pi)^3} \;\; \Pi_{\bar{A}\bar{D}}(\bfk_1) \Pi_{\bar{B}\bar{E}}(\bfp) \Pi_{\bar{C}\bar{F}}(\bfk_2 - \bfp) + {\mbox 5\, {\rm perm.}} \right] + \nonumber \\
&&  \frac{1}{2}N_{\bar{A}} N_{\bar{B}} N_{\bar{C}\bar{D}\bar{E}\bar{F}} \left[\int \frac{d^3p}{(2\pi)^3} \;\;\Pi_{\bar{A}\bar{C}}(\bfk_1) \Pi_{\bar{B}\bar{D}}(\bfk_{2} )  \Pi_{\bar{E}\bar{F}}(\bfp) + {\mbox 2\, {\rm perm.}} \right] + \nonumber \\
&& \frac{1}{2}N_{\bar{A}} N_{\bar{B}\bar{C}} N_{\bar{D}\bar{E}\bar{F}}\left[ \int \frac{d^3p}{(2\pi)^3}\;\;\Pi_{\bar{A}\bar{B}}(\bfk_1) \Pi_{\bar{C}\bar{D}}(\bfk_{12} )  \Pi_{\bar{E}\bar{F}}(\bfp) + {\mbox 5\, {\rm perm.}} \right] + \nonumber \\
&& \frac{1}{2} N_{\bar{A}} N_{\bar{B}} N_{\bar{C}\bar{D}} \left[\int \frac{d^3p}{(2\pi)^3}\;\;T_{\bar{A}\bar{B}\bar{C}\bar{D}}(\bfk_1, \bfk_2, \bfp, \bfk_3-\bfp ) + {\mbox 2\, {\rm perm.}} \right]  + \nonumber \\
&&   \frac{1}{2} N_{\bar{A}} N_{\bar{B}\bar{C}} N_{\bar{D}\bar{E}} \left[ \int \frac{d^3p}{(2\pi)^3} \;\; \Pi_{\bar{A}\bar{B}}(\bfk_1 ) B_{\bar{C}\bar{D}\bar{E}}(-\bfk_3, \bfp, \bfk_3 - \bfp ) + {\mbox 5\, {\rm perm.}} \right]  + \nonumber \\
&&   N_{\bar{A}} N_{\bar{B}\bar{C}} N_{\bar{D}\bar{E}} \left[ \int \frac{d^3p}{(2\pi)^3} \;\; B_{\bar{A}\bar{B}\bar{D}}(\bfk_1, \bfp, -\bfk_1 - \bfp)\Pi_{\bar{C}\bar{E}}(\bfk_2 - \bfp)  + {\mbox 2\, {\rm perm.}} \right] + \nonumber \\
&&  \frac{1}{2} N_{\bar{A}} N_{\bar{B}} N_{\bar{C}\bar{D}\bar{E}} \left[ \int \frac{d^3p}{(2\pi)^3} \;\; B_{\bar{A}\bar{B}\bar{C}}(\bfk_1, \bfk_2, \bfk_3)\Pi_{\bar{D}\bar{E}}(\bfp)  + {\mbox 2\, {\rm perm.}} \right] + \nonumber \\
&&  \frac{1}{2} N_{\bar{A}} N_{\bar{B}} N_{\bar{C}\bar{D}\bar{E}} \left[ \int \frac{d^3p}{(2\pi)^3} \;\; \Pi_{\bar{A}\bar{C}}(\bfk_1) B_{\bar{B}\bar{D}\bar{E}}(\bfk_2, \bfp, -\bfk_2 -\bfp ) + {\mbox 5\, {\rm perm.}} \right] .
\eea
From the equation above, we see that the terms from fifth to nineth carry the non-gaussian contributions encoded in the bispectrum $B_{\bar{A}\bar{B}\bar{C}}$ and the trispectrum $T_{\bar{A}\bar{B}\bar{C}\bar{D}}$ of the field perturbations. We can check the numerical factors in front of every term by invoking the rules number 7 and  number 8.
\section{Vertex renormalization}\label{ren_drvs}
\subsection{Gaussian fields}
\begin{figure}[h]
\centering
\includegraphics[scale=1.2]{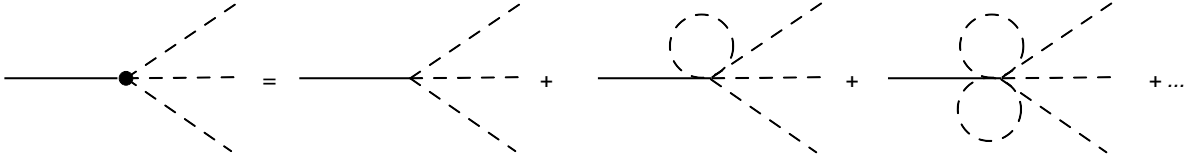}
\caption{The diagrammatic representation of the renormalized 3-vertex $\tilde{N}_{\bar{A} \bar{B} \bar{C}}$ for gaussian fields.}
\label{v_ren}
\end{figure}

\begin{figure}[h]
\centering
\includegraphics[scale=1.2]{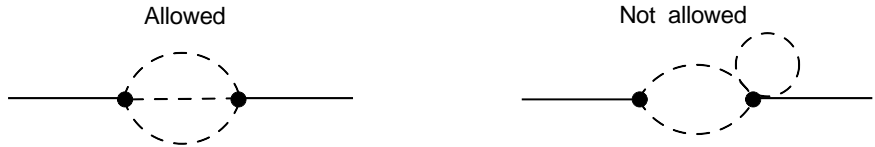}
\caption{An example of the type of diagrams allowed and not allowed in the perturbative expansion of $P_{\zeta}$.}
\label{yes-no}
\end{figure}

We can reduce significantly the number of diagrams by adopting the vertex renormalization procedure presented in Ref. \cite{Byrnes:2007tm}. While the discussion in Ref.  \cite{Byrnes:2007tm} is restricted to the context of multi-scalar fields perturbations, here, we extend the method to account for the case in which there are multi-scalar and multi-vector field perturbations. We describe here how this procedure works in the presence of gaussian fields and we leave the discussion of the non-gaussian case for next section. In this procedure the derivatives $N_{\bar{A} \bar{B} \bar{C} \cdots  }$ are redefined in such a way that they relate the values of $N$ at a given background $\Phi_{0}$  to the values of $N$ at a general point ${\vec x}$ in real space. The derivatives of $N$ for the background $\Phi_0$ are  $N_{\bar{A}\bar{B}\cdots }\vert_{\Phi_{0}} = N_{\bar{A}\bar{B}\cdots }$ and are related to the number of $e$-foldings $\tilde{N}$ at any point $\bfx$ by
\be\label{ntil}
\tilde{N} \equiv \tilde{N}(\Phi(\bfx)) = N + {N}_{\bar{A}}\delta\Phi_{\bar{A}} + \frac{1}{2}N_{\bar{A}\bar{B}}\delta\Phi_{\bar{A}}\delta\Phi_{\bar{B}}+ \frac{1}{3!}N_{\bar{A}\bar{B}\bar{C}}\delta\Phi_{\bar{A}}\delta\Phi_{\bar{B}}\delta\Phi_{\bar{C}}+ \cdots \,.
\ee
Taking the derivatives of this equation with respect to  $\Phi_{\bar{A}}$, then going to momentum space and taking the expectation value, we redefine the derivatives of $N$ in the form
\be\label{Ntilde}
\langle\tilde{N}_{\bar{A} \cdots \bar{B}}\rangle = N_{\bar{A} \cdots \bar{B}} + \frac{1}{2}N_{\bar{A}\cdots \bar{B} \bar{C} \bar{D}} \int \frac{d^3 p}{(2\pi)^3}  \Pi_{\bar{C} \bar{D}}({\bf{p}}) + \frac{1}{8}N_{\bar{A} \cdots \bar{B} \bar{C} \bar{D} \bar{E}\bar{F}} \int  \frac{d^3 p_{1}}{(2\pi)^3} \frac{d^3 p_{2}}{(2\pi)^3}  \Pi_{\bar{C} \bar{D}}({\bf{p}}_{1})\Pi_{\bar{E}\bar{F}}({\bf{p}}_{2}) + \cdots \,.
\ee
In the last step, we used the fact that the fields are gaussian, hence, all the expansion contains only integrals of propagators and not higher-order correlation functions. We see in Eq. (\ref{Ntilde}) that the vertex renormalization ``dresses" a vertex with an infinite series of the contractions of the vertices and the propagators integrated along independent internal momenta. Diagrammatically, this is equivalent to attach to the vertex an infinite series of loop integrals (``bubbles"). Graphically, we represent a ``dressed" vertex with a solid dot as shown in Fig. \ref{v_ren}. In this figure we show the expansion up to two-loops of the 3-vertex $\langle\tilde{N}_{\bar{A} \bar{B}\cdots \bar{C}}\rangle$. In order to simplify our notation, in the following we omit the brackets in the renormalized derivatives, then $\tilde{N}_{\bar{A} \bar{B}\cdots \bar{C}} \equiv \langle\tilde{N}_{\bar{A} \bar{B}\cdots \bar{C}}\rangle$.
When we introduce the renormalized vertex we just keep the diagrams present in the perturbative series of the correlation functions which do not have ``bubbles" around the vertices since all the ``bubbles" are already included in the renormalized vertex. In Fig. \ref{yes-no} we show an example of the type of diagrams which are allowed and the ones which are not allowed in the perturbative expansion of the power spectrum $P_{\zeta}$.

\subsection{Renormalized gaussian diagrams}
Taking into account the renormalization of the vertices, we can also write a set of rules for the renormalized diagrams as follows
\begin{enumerate}
\item Draw all distinct diagrams with $n$-external lines and $p$ propagators. 
External lines are represented by solid lines while internal lines are represented by dashed lines. 
Every renormalized vertex connect an external line to at least 1 propagator and is represented with a solid dot. The propagators cannot have both ends attached to the same renormalized vertex. See Fig. \ref{yes-no}.
\item Label the external lines with incoming momenta $\bfk_i$ for $i=1,...,n$ and label the propagator with internal momenta $\bfp_k$ for $k=1,...,p$. Label each end of each propagator with a field index: $\bar{A}, \bar{B}, \cdots, \bar{C}$. 
See Fig. \ref{prop}.
\item Assign the factor $\Pi_{\bar{A}\bar{B}}(\bfk)$ to each propagator in the diagram where $\bfk$ is the momentum associated to the propagator which must flow from the end with the label $\bar{A}$ to the end with the label $\bar {B}$ (see Fig. \ref{prop}). The direction of the momentum flow in the diagram is relevant when the action is not invariant under the parity transformation. If the momentum flows from the end with the label $\bar{B}$ to the end with the label $\bar{A}$, the assigned propagator must be $\Pi_{\bar{B}\bar{A}}(\bfk) = \Pi_{\bar{A}\bar{B}}(-\bfk)$.
\item Assign a factor $N_{\bar{A}_1\bar{A}_2\cdots \bar{A}_t} (2\pi)^3\delta(\bfk_i-\bfp_{1\cdots t})$ to each vertex, a $t$-vertex, where the number $t$ of derivatives of $N$ is the number of propagators attached to this vertex.  We use the convention that incoming momentum is positive. The Dirac delta function ensures that the momentum is conserved. See Fig. \ref{vert}.
\item Integrate over the propagator momenta $\frac{1}{(2\pi)^3}\int d^3p_i$. The first $n-1$ integrals can be done immediately using the Dirac delta functions but any further integral in general cannot be performed analytically. This is the case when there are integrals corresponding to loop corrections.
\item Divide by the appropriate numerical factor: $l!$ if there are $l$ propagators 
attached to the same vertices at both ends.
\item Add all permutations of the diagrams corresponding to all the distinct ways to relabel the $\bfk_i$ attached to the external lines. The number of permutations depends on the symmetries of the diagram: a diagram which is totally symmetric with respect to external lines has only one term; in contrast, a diagram without symmetries with respect to external lines has $n!$ permutations.
\end{enumerate}

As examples of these rules, we show the diagrams which represent the power spectrum up to two loops and the bispectrum up to one loop:
\bea
\langle\zeta(\bfk_!)\zeta(\bfk_2)\rangle^{\rm up\, to\, 2-loops}\ &=& \ \begin{array}{l}
\includegraphics[scale=0.77]{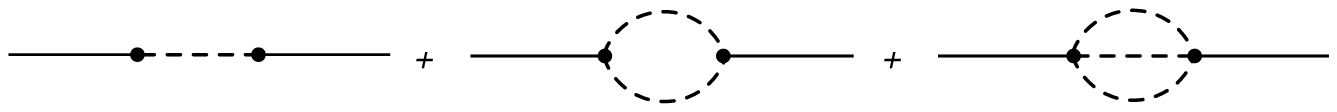} \,,
\end{array}\\
\langle\zeta(\bfk_1)\zeta(\bfk_2)\zeta(\bfk_3)\rangle^{\rm up\, to\, 1-loop}\ &=& \ \begin{array}{l}
\includegraphics[scale=0.77]{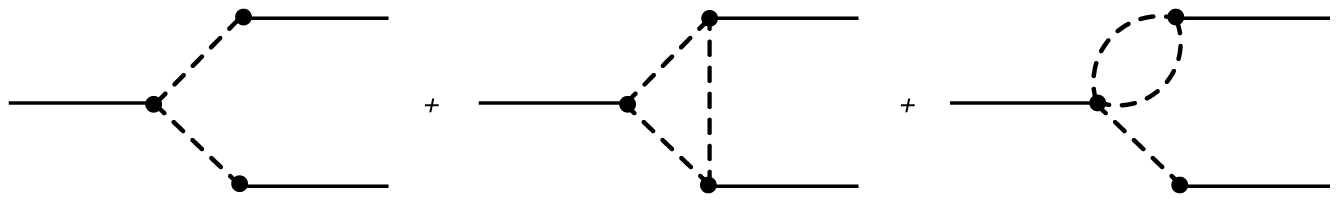} \,.
\end{array}
\eea
They represent the expansions
\bea \label{Pzeta_ren}
P_\zeta^{\rm (up\, to\, 2-loops)} ({\bfk_1}) &=&  \tilde{N}_{\bar{A}}\tilde{N}_{\bar{B}}\Pi_{\bar{A}\bar{B}}({\bfk_1})+ \frac{1}{2} \tilde{N}_{\bar{A}\bar{B}}\tilde{N}_{\bar{C}\bar{D}} \int \frac{d^3p}{(2\pi)^3}  \Pi_{\bar{A}\bar{C}}({\bfk_1}+\bfp)\Pi_{\bar{B}\bar{D}}(-\bfp) +  \nonumber \\
&&  \frac{1}{6}\tilde{N}_{\bar{A}\bar{B}\bar{C}} \tilde{N}_{\bar{D}\bar{E}\bar{F}} \int \frac{d^3p_1}{(2\pi)^3}  \frac{d^3p_2}{(2\pi)^3}  \Pi_{\bar{A}\bar{D}}(\bfp_1) \Pi_{\bar{B}\bar{E}}(\bfp_2) \Pi_{\bar{C}\bar{F}}({\bfk_1}-\bfp_{12}) \,, \\
\bz^{\rm (up\, to\, 1-loop)} (\bfk_1,\bfk_2,\bfk_3) &=& \tilde{N}_{\bar{A}} \tilde{N}_{\bar{B}} \tilde{N}_{\bar{C}\bar{D}}
\Big[ \Pi_{\bar{A}\bar{C}}(\bfk_1)\Pi_{\bar{B}\bar{D}}(\bfk_2) + 2 \ {\rm perm.}\Big]  + \no \\
&& \tilde{N}_{\bar{A}\bar{B}} \tilde{N}_{\bar{C}\bar{D}} \tilde{N}_{\bar{E}\bar{F}} \int \frac{d^3p}{(2\pi)^3} \;\; \Pi_{\bar{A}\bar{C}}(-\bfp) \Pi_{\bar{B}\bar{E}}(\bfk_1 + \bfp) \Pi_{\bar{D}\bar{F}}(\bfk_2 - \bfp) + \nonumber \\
&&  \frac{1}{2}\tilde{N}_{\bar{A}} \tilde{N}_{\bar{B}\bar{C}} \tilde{N}_{\bar{D}\bar{E}\bar{F}}\left[ \int \frac{d^3p}{(2\pi)^3} \;\; \Pi_{\bar{A}\bar{D}}(\bfk_1) \Pi_{\bar{B}\bar{E}}(\bfp) \Pi_{\bar{C}\bar{F}}(\bfk_2 - \bfp) + {\mbox 5\, {\rm perm.}} \right]  \,.
\eea
From the last equations, we see that the number of diagrams at each order reduces significantly, for instance the up to 2-loop power spectrum has only three diagrams instead of the seven diagrams that we have without renormalized vertices. In particular, the diagrammatic representation of the power spectrum is very simple because we have only one diagram at every order in loops. The $l$-loop diagram of the power spectrum is represented by $l+1$ propagators joining two $(l+1)$-renormalized vertices.
\subsection{Non-gaussian fields}
\begin{figure}[h]
\centering
\includegraphics[scale=0.95]{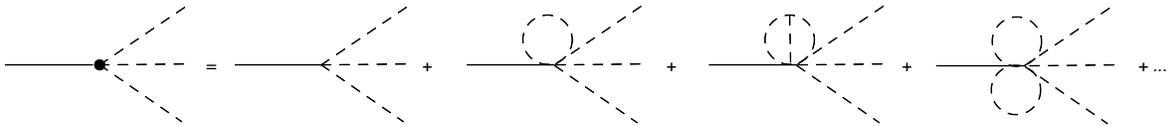}
\caption{The diagrammatic representation of the renormalized 3-vertex $\tilde{N}_{\bar{A} \bar{B} \bar{C}}$ for non-gaussian fields.}
\label{v_ren_ng}
\end{figure}

We start again from Eq. (\ref{ntil}), we take the derivatives of this equation with respect to $\Phi_{\bar{A}}$, then go to momentum space, and finally take the expectation value. Thus, we redefine the derivatives of $N$ in the form
\bea\label{NtildeNG} \nonumber
\langle\tilde{N}_{\bar{A} \cdots \bar{B}}\rangle &=& N_{\bar{A} \cdots \bar{B}} + \frac{1}{2}N_{\bar{A}\cdots \bar{B} \bar{C} \bar{D}} \int \frac{d^3 p}{(2\pi)^3}  \Pi_{\bar{C} \bar{D}}({\bf{p}}) + \frac{1}{3!}N_{\bar{A} \cdots \bar{B} \bar{C} \bar{D} \bar{E}} \int  \frac{d^3 p_{1}}{(2\pi)^3} \frac{d^3 p_{2}}{(2\pi)^3}  B_{\bar{C} \bar{D}\bar{E}}({\bf{p}}_{1}, -{\bf{p}}_{2}, {\bf{p}}_{2}-{\bf{p}}_{1}) + \\
&& \frac{1}{8}N_{\bar{A} \cdots \bar{B} \bar{C} \bar{D} \bar{E}\bar{F}} \int  \frac{d^3 p_{1}}{(2\pi)^3} \frac{d^3 p_{2}}{(2\pi)^3}  \Pi_{\bar{C} \bar{D}}({\bf{p}}_{1})\Pi_{\bar{E}\bar{F}}({\bf{p}}_{2}) + \cdots \,.
\eea
In this case, we consider that the fields are non-gaussian so that  the integrals in the expansion contain propagators and higher order correlation functions as well. In Eq. (\ref{NtildeNG}) we have now that the renormalized vertices will contain not only propagator loops but in general they will have $(k-1)$-spectra with all their lines wrapped around the vertices. Graphically, we represent a ``dressed" vertex with non-gaussian fields with a solid dot as shown in Fig. \ref{v_ren_ng}. In this figure we show the expansion up to two-loops of the 3-vertex $\tilde{N}_{\bar{A} \bar{B}\cdots \bar{C}}$. We see for instance in the third term of such a figure that there is a 2-spectrum wrapping around the vertex.

The diagrammatic rules for this situation are the same as the ones listed in Section \ref{ngdrvs}, but excluding the diagrams containing $(k-1)$-spectra with all their lines attached to the same vertex. A general proof that the redefinition of the derivatives given in Eq. (\ref{NtildeNG}) remove all the diagrams with dressed vertices was given in the appendix of Ref. \cite{Byrnes:2007tm}.
\section{Conclusions}     
Formulating the $n$-point correlators of the primordial curvature perturbation $\zeta$ in terms of integrals involving correlators of the field perturbations is a very direct procedure; however it is in general very time-consuming, clutter, and not so intuitive.  The same situation happened with the calculation of scattering amplitudes in cannonical Quantum Field Theory until Feynman exposed for the first time his diagrammatic rules \cite{feynman1,feynman2,feynman3} which, since then, have become the standard way of doing such a kind of calculations \cite{diagrammatica}; the reason: Feynman rules are vey easy to remember, very easy to implement, very intuitive, and allow us to understand the physics behind a quantum process.  A few years ago, the ``Feynman-like rules'' to calculate $n$-point correlators of $\zeta$ were developed \cite{Byrnes:2007tm}, with the same aims as the Feynman rules in the particle physics context.  In that work, only scalar fields were introduced as the generators of $\zeta$.  Since then, not many cosmologists have used this diagrammatic approach, wasting the interesting properties stated above.  As mentioned in the introduction, the relevance of having an efficient method to calculate loop corrections relies mainly in the possibility that observational data coming from high precision cosmological probes could open a window to access effects related to such corrections; there are also scenarios in which evaluating loop contributions is essential because those terms dominate over the tree level ones and constitute a source for large and observable non-gaussianities \cite{vrl, vr, Boubekeur:2005fj, Cogollo:2008bi, Rodriguez:2008hy, Kumar:2009ge}.  In the present work, we have been interested in promoting even more this methodology among the cosmology community. In addition, we have extended it to the interesting case where multiple scalar and vector fields contribute to the generation of $\zeta$; in this situation it is possible to obtain prolonged stages of anisotropic expansion (see for instance Ref. \cite{watanabe1,anexp1,anexp2,anexp3}), as well as observable levels of statistical anisotropy \cite{dklr,abramo,bartolo3} and anisotropic non-gaussianity \cite{dimopoulos,mythesis,bartolo3} which can serve as discriminators among different inflationary models \cite{wmap7,groeneboom}.  Indeed, consistency relations among the different levels of non-gaussianity and the level of statistical anisotropy may be obtained for different classes of models so that they may easily be ruled out by observation \cite{vrl,vr,mythesis,beltran}.  Two of the most notorious differences with respect to the multi-scalar field case of Ref. \cite{Byrnes:2007tm} are the possibility of parity violating interactions in the action and the possibility of anisotropic expansion; in the former case, the direction of the momentum flow in the Feynman-like diagrams is relevant, something which is not present for scalar and vector boson propagators in the Feynman rules of the particle physics context;  in the latter, the particle production process is statistically anisotropic, which render  the scalar-scalar field perturbation spectra and the vector-vector field perturbation spectra for the different polarizations wavevector-dependent, as well as making the correlators between scalar and vector field perturbations different to zero (producing scalar-vector field perturbation spectra).  In view of recent relevant works where the anisotropic expansion is quite prolonged \cite{himmetoglu4,watanabe2,dulaney,peloso,watanabe1}, we are urged to comprehensively study the generation of statistical anisotropy in that kind of models. 

\section*{Acknowledgments}
C.A.V.-T. is supported by Vicerrector\'{\i}a de Investigaciones (UNIVALLE) grant number 7858. Y.R. is supported by DIEF de Ciencias (UIS) grant number 5177; he also acknowledges the hospitality of the YITP - Yukawa Institute for Theoretical Physics at Kyoto University (Japan) during the workshop YITP-T-10-01 {\it ``Gravity and Cosmology 2010''}. J.P.B.A. is supported by VCTI (UAN) under project 2010251 and is grateful
to the Dipartimento di Fisica ``Galileo Galilei'' at Universit\`{a} degli
Studi  di Padova (Italy) for kind hospitality during the completion of this
work.

\renewcommand{\refname}{{\large References}}

\end{document}